\documentclass[12pt] {article}		
\usepackage{epsfig}

\normalsize
\newcommand{\be}{\begin{equation}}
\newcommand{\ee}{\end{equation}}
\newcommand{\bea}{\begin{eqnarray}}
\newcommand{\eea}{\end{eqnarray}}
\newcommand{\gsim}{\hbox{ \raise3pt\hbox to 0pt{$>$}\raise-3pt\hbox{$\sim$} }}
\newcommand{\lsim}{\hbox{ \raise3pt\hbox to 0pt{$<$}\raise-3pt\hbox{$\sim$} }}
\newcommand{\mathbold}[1]{\mbox{\boldmath $\bf#1$}}
\newcommand{\figdir}{figs}
\begin{document}

\begin{titlepage}
\hfill
\vskip 4.0cm
\begin{center}
\large \bf
Development of a Geant4 Solid for Stereo Mini-jet Cells \\
in a Cylindrical Drift Chamber
\end{center}

\vskip 1.5cm

\begin{center}
\large 
$Kotoyo$ $Hoshina^a$, $Keisuke$ $Fujii^{a}$\footnote{
Corresponding authhor.\\
E-Mail address: fujiik@jlcuxf.kek.jp\\
TEL: +8-298-64-5373\\
FAX: +8-298-64-2580
}, and $Osamu$ $Nitoh^b$,
\end{center}

\vskip 0.7cm

\begin{center}
$^a$ High Energy Accelerator Research Organization(KEK),
 Tsukuba, 305-0801, Japan\\ 
$^b$Tokyo University of Agriculture and Technology,
 Tokyo 184-8588, Japan\\
\end{center}

\vskip 1cm

\begin{abstract}
Stereo mini-jet cells will be indispensable components 
of a future $e^+e^-$ linear collider central tracker such as JLC-CDC.
There is, however, no official Geant4 solid available
at present to describe such geometrical objects, 
which had been a major obstacle for us
to develop a full Geant4-based simulator with stereo cells
built in.
We have thus extended Geant4 to include a new solid 
({\tt TwistedTubs}), which consists of three kinds of
surfaces: two end planes, inner and outer hyperboloidal
surfaces, and two so-called twisted surfaces that make
slant and twisted $\phi$-boundaries.
Design philosophy and its realization in the Geant4 framework 
are described together with algorithmic details.
We have implemented stereo cells with the new solid,
and tested them using geantinos and
Pythia events ($e^{+}e^{-}\rightarrow ZH$ at $\sqrt{s} = 350~{\rm GeV}$).
The performance was found reasonable:
the stereo cells consumed only 25\% more CPU time than
ordinary axial cells.
\end{abstract}
\vfil

\end{titlepage}


\section{Introduction}

Experiments at a future linear $e^+e^-$ collider such as JLC\cite{Ref:JLC-I}
will open up a novel possibility to reconstruct all the final states
in terms of fundamental particles (leptons, quarks, and gauge bosons).
This involves identification of heavy unstable particles
such as $W$, $Z$, and $t$ through jet invariant-mass measurements.
High resolution energy flow measurements will thus be crucial,
necessitating high resolution tracking and calorimetry
as well as good track-cluster matching
to avoid double counting.
A large cylindrical drift chamber with small jet cells 
(JLC-CDC\cite{Ref:acfa}) is
our choice for a candidate central tracking device to
fulfill these requirements.
Good track-cluster matching requires, however,
small track extrapolation errors,
which in turn demand
high $r$-$\phi$ and $z$ resolutions.
The latter forces us to
introduce cells consisting of stereo wires, 
since the $z$ resolution
in charge division or time difference readout
is typically $1~\%$ of the wire length or worse,
which is a few cm in the linear collider use.

We have already studied and published 
hardware aspects of common problems 
in designing stereo-wire geometry
for a long cylindrical drift chamber 
with small jet cells\cite{Ref:stereo-gain}.
In order to finalize the chamber design so as to
achieve the best attainable
energy flow resolution, however, we
need to carefully optimize
the layout of axial and stereo cells
through detailed Monte Carlo simulations.
Considering the recent advance of
object-oriented technology in high energy physics
software development,
we have thus started the development of a full
Monte Carlo simulator called JUPITER\cite{Ref:Jupiter}
based on Geant4\cite{Ref:geant4toolkit}.

The axial layers were easy to implement, using a 
standard Geant4 solid called {\tt G4Tubs},
which is a $\phi$ segment of a cylinder.
On the other hand, there is currently no official
Geant4 solid available to describe stereo cells,
which had been a major obstacle for us to
install stereo cells into the full simulator. 
We have thus extended Geant4 to include a new
solid ({\tt TwistedTubs}), which comprises
three kinds of bounding surfaces:
two end planes, inner and outer hyperboloidal surfaces,
and two  so-called twisted surfaces that make
slant and twisted $\phi$ boundaries.
Although {\tt TwistedTubs} was developed
under the JUPITER environment,
it is actually a general purpose Geant4 extension.
This paper describes the design philosophy
of {\tt TwistedTubs},
its realization in the Geant4 framework, 
and algorithmic details,
together with results of its performance test.

The paper is organized as follows.
We begin with a brief account of geometrical parameters
that determine the configuration of a stereo cell,
and then review the basic procedure to add a new solid
to Geant4.
The subsequent two sections are devoted to descriptions
of design philosophy, implementations, and algorithmic details
of {\tt TwistedTubs},
which is followed by presentation of test results and discussions.
Finally section \ref{Sec:conclusions} summarizes our
achievement and concludes this paper.


\section{Geometry of Stereo Mini-jet Cell}
\label{Sec:geometry}

\begin{figure}[htb]
\begin{minipage}[htb]{7cm}
\centerline{
\epsfysize=6cm 
\epsfbox{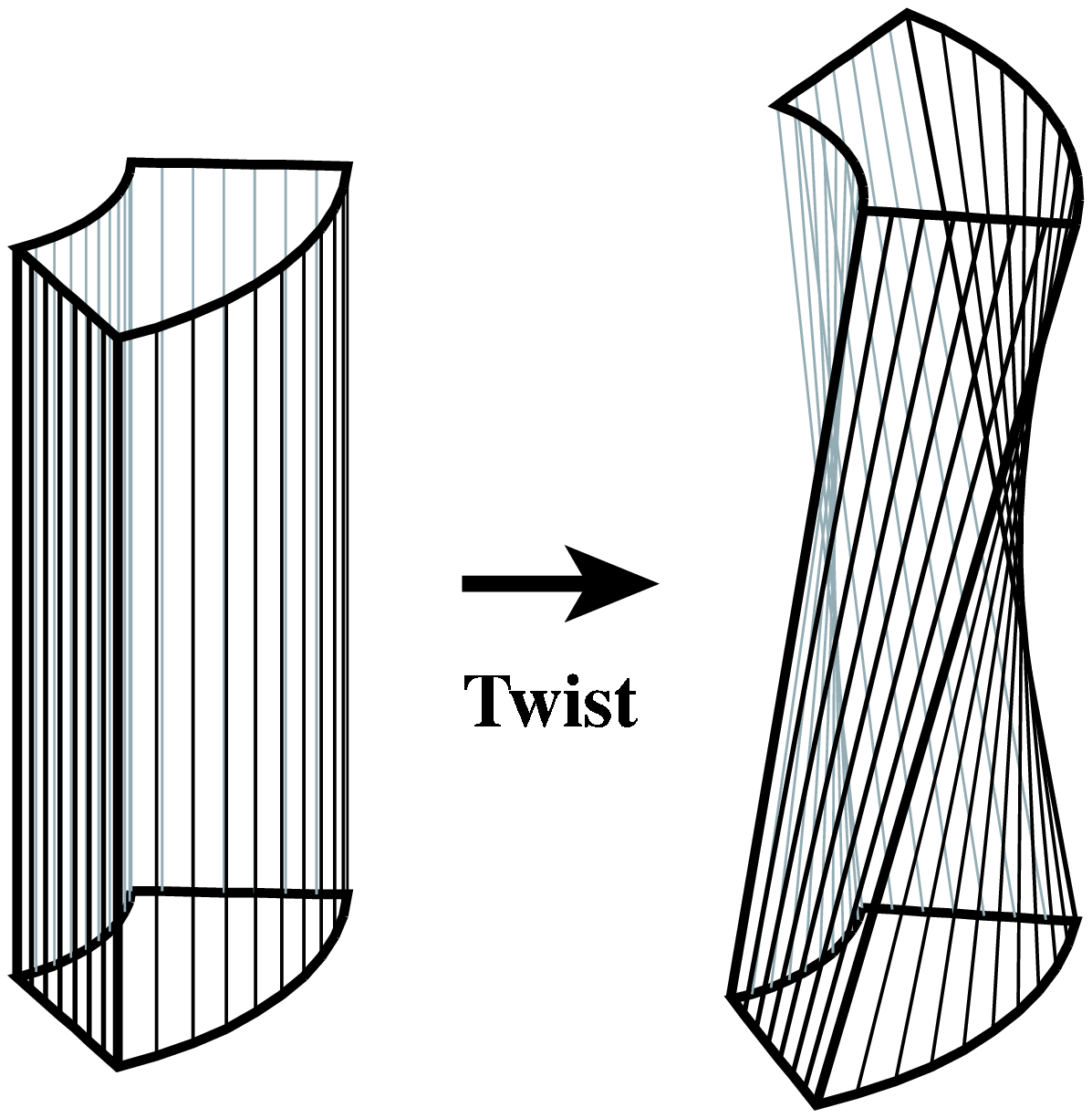}
}
\caption[Fig:twistedtubs]{\small \label{Fig:twistedtubs}
An exaggerated illustration of a stereo cell 
as formed by twisting an axial cell.
}
\end{minipage}
\hfill
\begin{minipage}[htb]{7cm}
\centerline{
\epsfysize=6cm 
\epsfbox{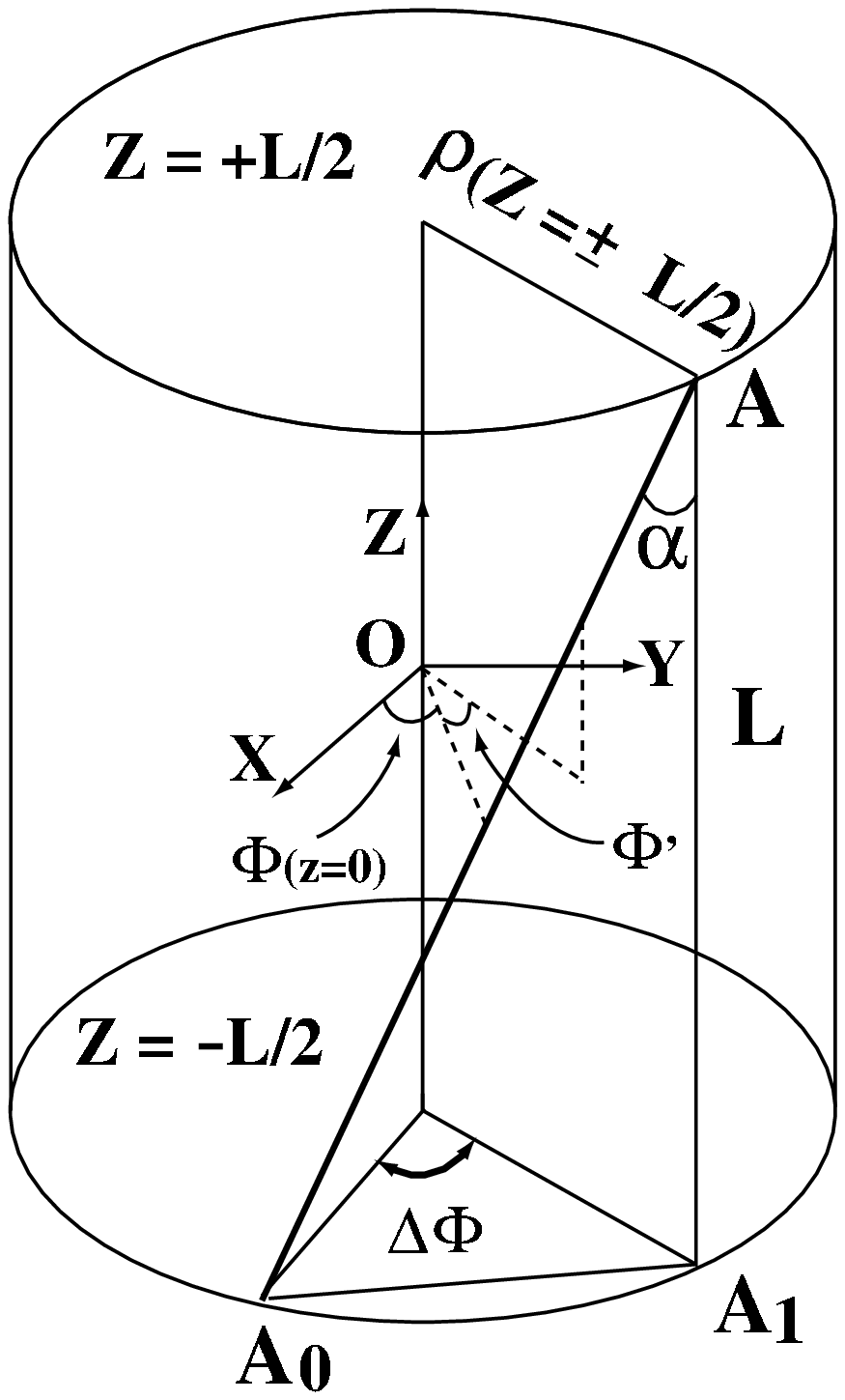}
}
\caption[Fig:3d-stereo-wire]{\small \label{Fig:3d-stereo-wire}
		3-dimensional view of a single stereo wire.
}
\end{minipage}
\end{figure}

In this section
we introduce, following the convention of
\cite{Ref:stereo-gain}, the stereo-geometrical parameters
that will be needed in subsequent sections.

Consider a cylindrical tube
consisting of two cylindrical layers of axial field-shaping wires
strung across two disc-shaped end plates
at some radii, $\rho_{in}$ and $\rho_{out}$.
An axial mini-jet cell is a segment cut out from this cylindrical tube
by $\phi$ boundaries formed by cathode wires.
As illustrated in Fig.~\ref{Fig:twistedtubs},
twisting one end plate
by a twist angle $\Delta \phi$
with the other end fixed
turns the axial mini-jet cell into a stereo mini-jet cell.
Inspection of the figure tells us that the stereo cell consists of
three pairs of three kinds of surfaces:
a pair of hyperboloidal surfaces setting inner and outer radial boundaries,
another pair of so-called twisted surfaces making left and right azimuthal boundaries,
and yet another pair of fan-shaped flat surfaces closing the positive and negative
$z$ ends of the cell.

By construction,
any of the four side walls of the stereo cell can be
regarded as a locus of a stereo wire sweeping through
the surface.
The geometry of a given side surface can thus be completely
determined by the equation for a single representative wire
chosen from the stereo wires forming that surface.
As depicted in Fig.~\ref{Fig:3d-stereo-wire},
the representative stereo wire is uniquely specified by
the radius at the ends $\rho(z=\pm L/2)$ or that at the
center $\rho_c \equiv \rho(z=0)$, the projected wire length $L$
to the chamber axis, 
and the twist angle $\Delta \phi$.
Notice that $\Delta \phi$ is signed and measured from $A_0$ to $A_1$.
The stereo angle $\alpha$, which is defined as an angle between
$AA_0$ and $AA_1$, is also signed, 
having the same sign as $\Delta \phi$.

\begin{figure}
\begin{center}
\begin{minipage}[htb]{6cm}
\centerline{
\epsfysize=9cm
\epsfbox{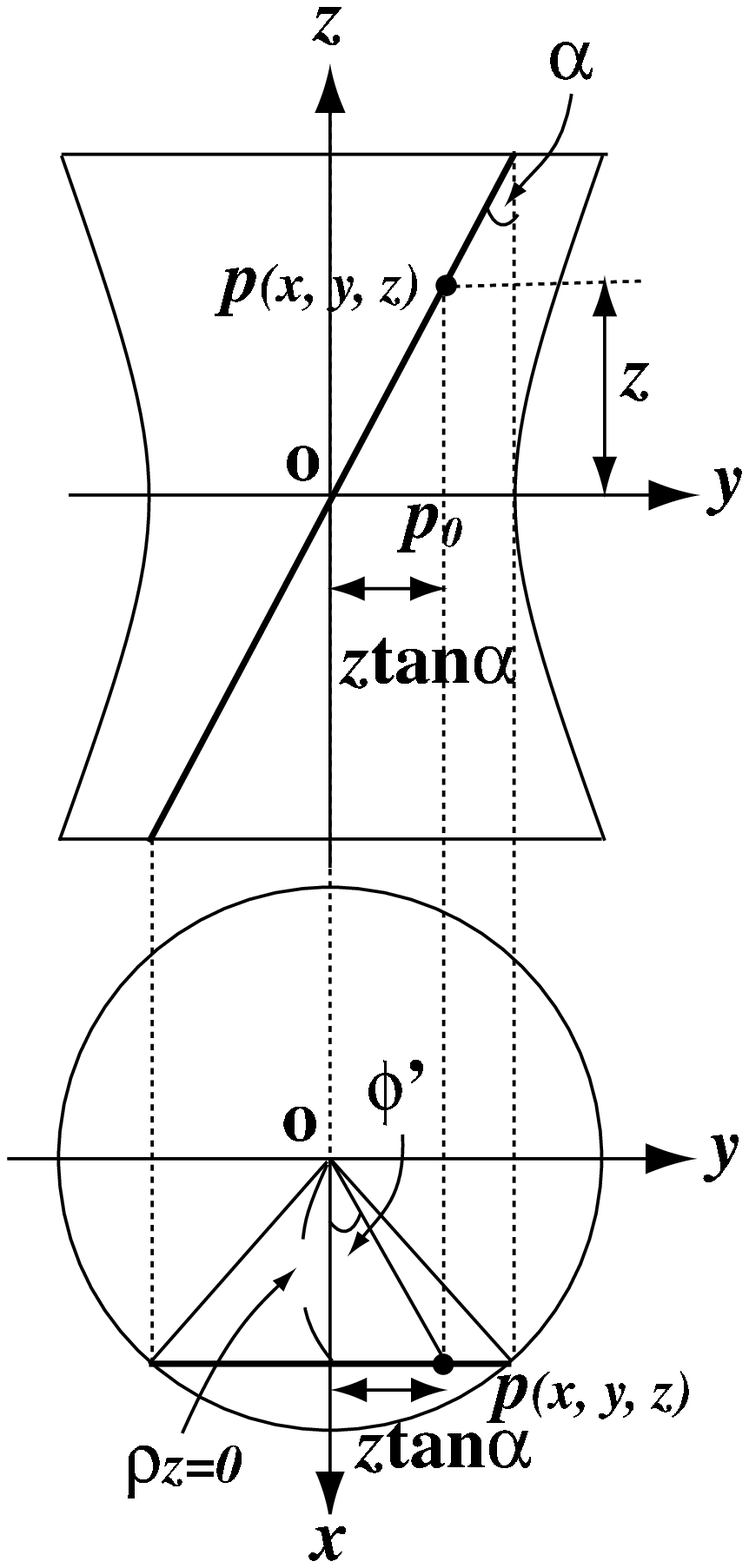}}
\caption[Fig:HyperboloidalSurface]{\label{Fig:HyperboloidalSurface}
Hyperboloidal surface and a typical line (wire) contained in it.
}
\end{minipage}
\hspace*{0.5cm}
\begin{minipage}[htb]{6cm}
\centerline{
\epsfysize=9cm
\epsfbox{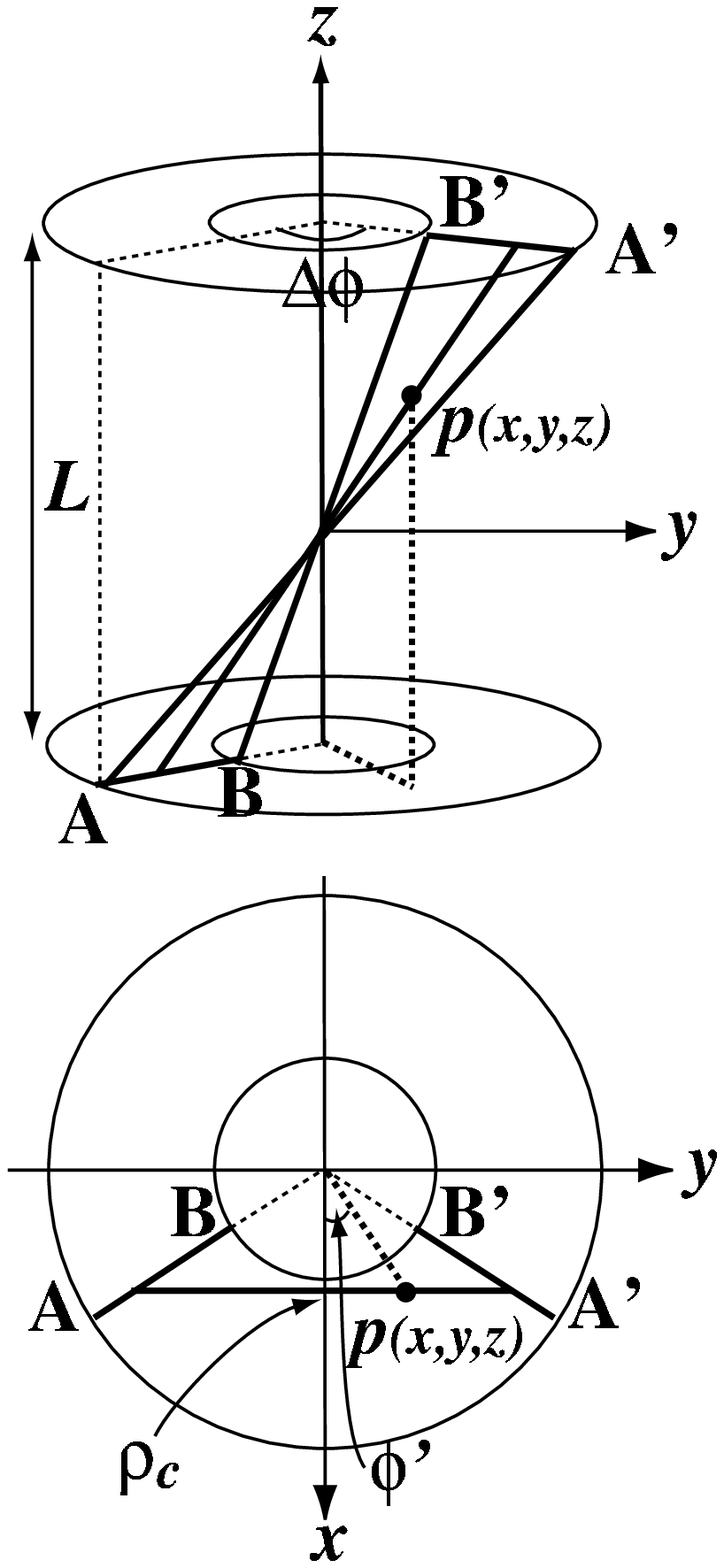}}
\caption[Fig:TwistedSurface]{\label{Fig:TwistedSurface}
Similar picture to Fig.~\ref{Fig:HyperboloidalSurface} for
a twisted surface.
}
\end{minipage}
\end{center}
\end{figure}

The stereo angle $\alpha$ can now be written in terms of 
$\rho(z=\pm L/2)$, $\Delta\phi$,
and the projected wire length ($L$):
\begin{equation}
   \label{Eq:alpha(r)}
   { \everymath{\displaystyle}
   \begin{array}{lll}
   	\alpha & = & \tan^{-1}\left( \frac{2 \rho(z=\pm L/2)}{L} 
   		\sin\left(\frac{\Delta\phi}{2}\right) \right) 
   \end{array}
   }
\end{equation}
It is obvious from Fig.~\ref{Fig:3d-stereo-wire} that
both the azimuthal angle and the radial position
of the stereo wire become $z$-dependent:
\begin{eqnarray}
\label{Eq:phi(z)}
   \phi(z) & = & \phi(z=0) + \phi ' \cr
	      & = & \phi(z=0) + \tan^{-1} \left[ \left(\frac{2z}{L}\right)
                               \tan\left(\frac{\Delta\phi}{2}\right) \right] \\
\rule{0in}{5.0ex}
\label{Eq:r(z)}
  \rho(z) & = & \sqrt{ \left( \rho(z=0) \right)^2 + \left( z\tan\alpha \right)^2 } \cr
\rule{0in}{4.0ex}
     ~ & = & \sqrt{ \left( \rho(z=\pm L/2) \right)^2 + \left( z^2 - (L/2)^2 \right) \tan^2\alpha},
\end{eqnarray}
where $z$ is measured from the middle of the chamber along the
chamber axis as Fig.~\ref{Fig:HyperboloidalSurface} indicates.
\\

Now we consider a hyperboloidal surface obtained as a locus of the straight line
given by Eqs.~\ref{Eq:phi(z)} and \ref{Eq:r(z)} by sweeping $\phi(z=0)$
from $\phi_{left}$ to $\phi_{right}$ with $\rho(z=0)$ fixed.
Let the outward normal to the hyperboloidal surface at a surface point
$\mathbold{x} = (x, y, z)$ be $\mathbold{n}$.
Apparently the outward normal $\mathbold{n}$ is
in the $\rho$-$z$ plane containing $\mathbold{x}$
and is perpendicular to the tangential vector thereat.
Eq.~\ref{Eq:r(z)} tells us that the cross section of the hyperboloidal surface
by  the $\rho$-$z$ plane becomes a hyperbola given by 
\begin{eqnarray*}
\left\{
\begin{array}{rcl}
\rho &=& \sqrt{\left(\rho(z=0)\right)^2 + \left( z \tan \alpha \right)^2} 
 \cr \rule{0in}{3.0ex}
z &=& z.
\end{array}
\right.
\end{eqnarray*}
The tangential vector we need is thus obtained by differentiating this equation
with respect to $z$.
Normalizing it to $\left| \mathbold{x} \right|$, we obtain
\begin{eqnarray}
\mbox{the tangential vector} &=& (z \tan^2 \alpha, ~\rho),
\label{Eq:HypeTangential}
\end{eqnarray}
\noindent
which leads to the outward normal in the $\rho$-$z$ plane:
$\mathbold{n}_\rho = (\rho, - z \tan^2\alpha)$.
Recalling that $\rho$ in the $\rho$-$z$ plane corresponds to $(x, y)$
in the $x$-$y$ plane,
we now obtain
\begin{equation}
\mathbold{n} = (x, y, -z~\tan^2 \alpha).
\label{Eq:normhype2}
\end{equation}
\noindent
The inward normal is anti-parallel with this and can be
obtained by simply changing the signs of all the components.
\\

On the contrary to the hyperboloidal surface, 
a twisted surface is a locus of the straight line
given by Eqs.~\ref{Eq:phi(z)} and \ref{Eq:r(z)}, when 
$\rho_c \equiv \rho(z=0)$ is swept from 
$\rho_{c, in} \equiv \sqrt{\rho_{in}^2 - \left((L/2)\tan\alpha\right)^2}$
to $\rho_{c, out} \equiv \sqrt{\rho_{out}^2 - \left((L/2)\tan\alpha\right)^2}$
with $\phi(z=0)$ fixed, 
where $\rho_{in}$ and $\rho_{out}$ are the inner and outer radii at the end planes
as defined before.
By construction,
any point on such a twisted surface
can be specified by two parameters ($\rho_c$ and $z$) 
in a local coordinate system for which
$x$-axis is chosen such that $\phi(z=0) = 0$,
as depicted in Fig.~\ref{Fig:TwistedSurface}:
\begin{eqnarray}
\left\{
\begin{array}{rclcl}
x & = & \rho_c \cr
y & = & \rho_c \tan \phi'  & = &  \rho_c \, \kappa \, z \cr
z & = & z,
\end{array}
\right.
\label{Eq:Twisted}
\end{eqnarray}
\noindent
where 
$\tan\phi '$
is given, through Eq.~\ref{Eq:phi(z)}, by
\begin{eqnarray}
\tan\phi' & = & \frac{2z}{L} \cdot \tan\left(\frac{\Delta \phi}{2}\right)
\end{eqnarray}
while $\kappa$ is defined to be
\begin{eqnarray}
\kappa & \equiv &  \frac{2}{L} \cdot \tan\left(\frac{\Delta \phi}{2}\right).
\end{eqnarray}
Notice that the $\Delta \phi$ must be smaller than $\pi$, implying
that $\rho_c > 0$ 
in this coordinate system.
Eq.~\ref{Eq:Twisted} shows that, in addition to the lines corresponding to cathode wires,
the twisted surface contains another kind of
straight lines that appear as the sections of different $z$-slices of the twisted surface.
Each of these lines corresponds to a $\phi$ boundary of a $z$-slice of a stereo cell
and passes through the $z$-axis.
At the point \mathbold{x} on the surface,
a pair of these two kinds of straight lines passing through it
divides the tangential directions at that point into four regions: 
the surface turns from convex to concave or vice versa as crossing
the boundaries.
In other words, the twisted surface has a saddle shape\footnote{
It is well known that the hyperboloidal surface has also a saddle shape.
In this case, the two straight lines over which the curvature changes
its sign are in the directions of stereo angles of $\pm \alpha$.
}.

The two kinds of straight lines contained in the twisted surface 
are useful when we construct a normal
to the twisted surface as we shall see below.
In general a normal to the surface at a given surface point 
$\mathbold{x} = (x, y, z)$
can be formed as a vector product
of two linearly independent tangential vectors at that point.
In our case such tangential vectors can be readily obtained by
partially differentiating
Eq.~\ref{Eq:Twisted} with respect to $x = \rho_c$ and $z$:
\begin{eqnarray} 
\begin{array}{@{\, }llllc@{\,}} 
{\bf e}_z & \equiv & \left(
                     \begin{array}{@{\, }c@{\,}}
					       \frac{\partial{x}}{\partial{z}} \cr
						   \frac{\partial{y}}{\partial{z}} \cr
						   \frac{\partial{z}}{\partial{z}}  
				     \end{array}
					 \right)  
 & = & \left(
                     \begin{array}{@{\, }c@{\,}}
					        0 \cr
						    \rho_c \kappa \cr
						    1  
				     \end{array}
					 \right) \cr
					     \rule{0in}{7.0ex}
{\bf e}_x & \equiv & \left(
                     \begin{array}{@{\, }c@{\,}}
					       \frac{\partial{x}}{\partial{x}}  \cr
						   \frac{\partial{y}}{\partial{x}}  \cr
						   \frac{\partial{z}}{\partial{x}}  
				     \end{array}
					 \right) 
& = & \left(
                     \begin{array}{@{\, }c@{\,}}
					       1 \cr
						   \kappa z \cr
						   0  
				     \end{array}
					 \right).
\end{array}
\label{Eq:TwistedNormal}
\end{eqnarray}
The first of the above two tangential vectors, ${\bf e}_z$, 
can virtually be identified as one of the cathode wires
forming the twisted surface.
On the other hand, the second one, ${\bf e}_x$ can be
viewed as a $\phi$ boundary of a $z$-slice of a stereo cell.
Using Eq.~\ref{Eq:TwistedNormal}, we can now easily form 
the surface normal at the surface point $\mathbold{x}$
by taking the vector product of these two tangential vectors.
\\

Armed with the equations given in this section,
we shall, in the next two sections,
prepare classes 
to represent the three kinds of surfaces:
{\tt J4HyperboloidalSurface}s, {\tt J4TwistedSurface}s, and {\tt J4FlatSurface}s,
and assemble them to form 
a Geant4 solid class ({\tt J4TwistedTubs})
to describe our stereo mini-jet cells of JLC-CDC.

\section{How to Add a New Solid to Geant4}

According to Geant4 User's Guide for Toolkit Developers\cite{Ref:geant4toolkit},
every Geant4 solid has to inherit from a base class called {\tt G4VSolid}.
This base class has the following pure virtual functions:
\begin{flushleft}
\underline{\tt G4VSolid}
\end{flushleft}
\begin{description}
 	\item {\tt G4double DistanceToIn(const G4ThreeVector \&p)} \\
	 			to calculate the minimal (or shorter) distance to the solid from an outside point ({\tt p}),
 	\item {\tt G4double DistanceToIn(const G4ThreeVector \&p, const G4ThreeVector \&v) }\\
	 			to calculate the exact distance from the outside point {\tt p} 
				 in the direction of a velocity vector {\tt v} to the solid,
	\item {\tt G4double DistanceToOut(const G4ThreeVector \&p)} \\
	 			to calculate the minimal (or shorter) distance to the solid from an inside point {\tt p},
	\item {\tt G4double DistanceToOut(const G4ThreeVector \&p, const G4ThreeVector \&v, \\
			const G4bool calcNorm=FALSE, G4bool *validNorm=0, G4ThreeVector *n) }\\
				to calculate the exact distance from the inside point {\tt p} to the solid along a velocity vector {\tt v},
	\item {\tt G4ThreeVector SurfaceNormal(const G4ThreeVector \&p)} \\
				which returns the outward unit normal at a surface point {\tt p}  
				(or, if {\tt p} is not on the surface, at the surface point that is the closest from {\tt p}),
	\item {\tt EInside Inside(const G4ThreeVector \&p) }\\
				to judge whether a space point {\tt p} is inside, or outside, 
				or on the surface of the solid and returns kInside, or kOutside, or kSurface, accordingly,
	\item {\tt G4bool CalculateExtent(const EAxis pAxis, const G4VoxelLimits \&pVoxelLimit,
			const G4AffineTransform \&pTransform, G4double \&pMin,
			G4double \&pMax) const }\\
				to calculate minimum and maximum extents ({\tt pMin} and {\tt pMax}) of the solid 
				in the direction of a given coordinate axis {\tt pAxis} within the limits specified by 
				{\tt pVoxelLimit} under an Affine transformation {\tt pTransform}, and
	\item {\tt G4GeometryType GetEntityType() const} \\
				to identify geometry type of the solid (necessary for persistency and STEP interface, but otherwise unused).
\end{description}
\noindent
The first five involve surfaces that bound the solid and can be calculated
on a surface-by-surface basis.
On the other hand the remaining three require information of the solid as a whole.
Any new user defined solid thus has to be equipped with these functions. 
Geant4 also provides BREPS\footnote{Boundary REPresented Solid. } classes for the purpose of 
constructing a solid from boundary surfaces. 
BREPS are, however, primarily designed to facilitate implementation 
of a solid with a complicated shape through interface like STEP to a CAD system. 
Since the stereo cell geometry can be handled analytically as explained in the
subsequent sections, 
we decided to develop a new dedicated solid class for it.

\section{Design of J4TwistedTubs}

\begin{figure}
\begin{center}
\begin{minipage}[htb]{8cm}
\centerline{
\epsfysize=8cm
\epsfbox{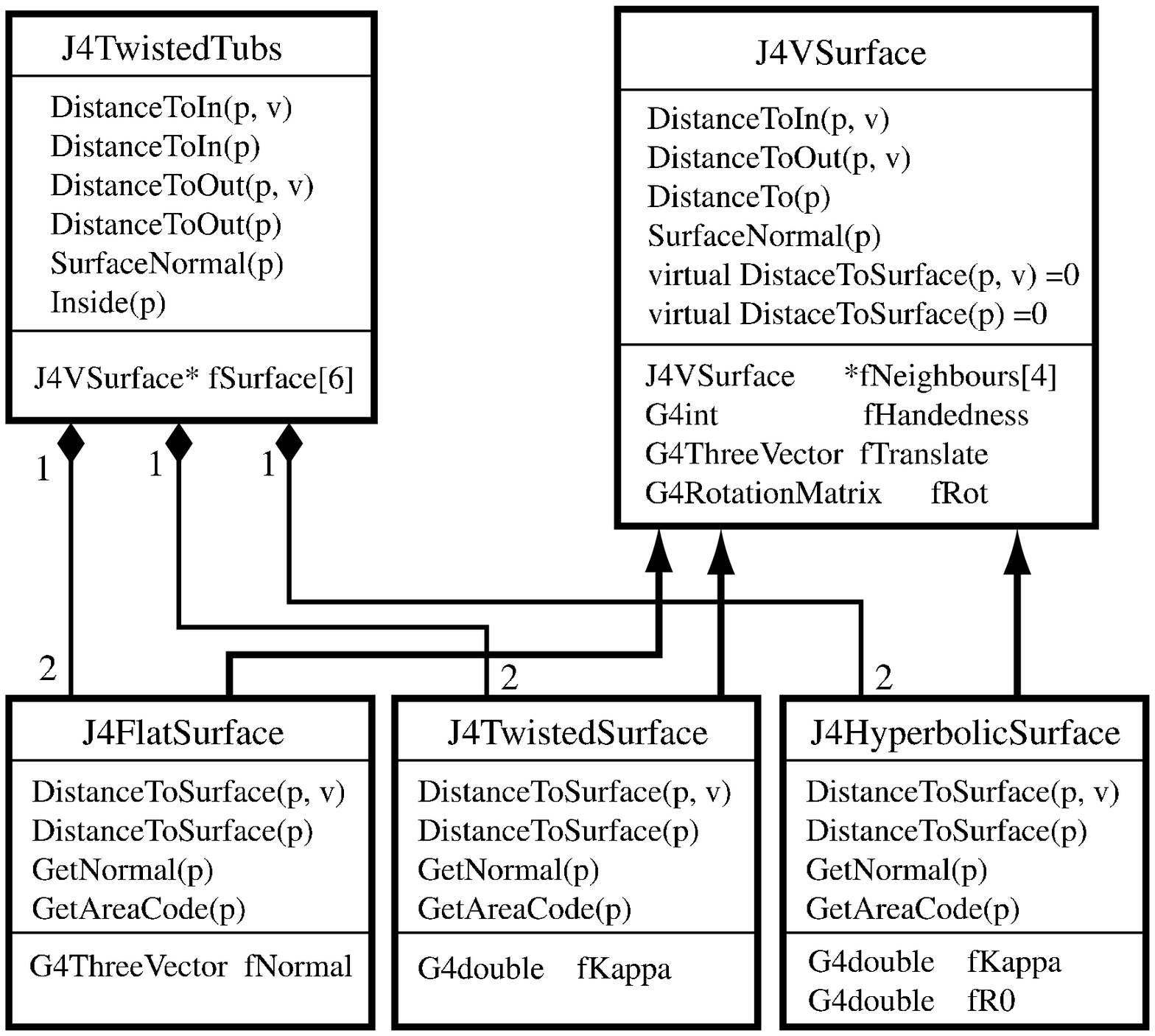}}
\caption[Fig:TwistedTubsUML]{\label{Fig:TwistedTubsUML}
A class diagram in UML of {\tt J4TwistedTubs}
}
\end{minipage}
\hspace*{1cm}
\begin{minipage}[htb]{6cm}
\centerline{
\epsfysize=7cm
\epsfbox{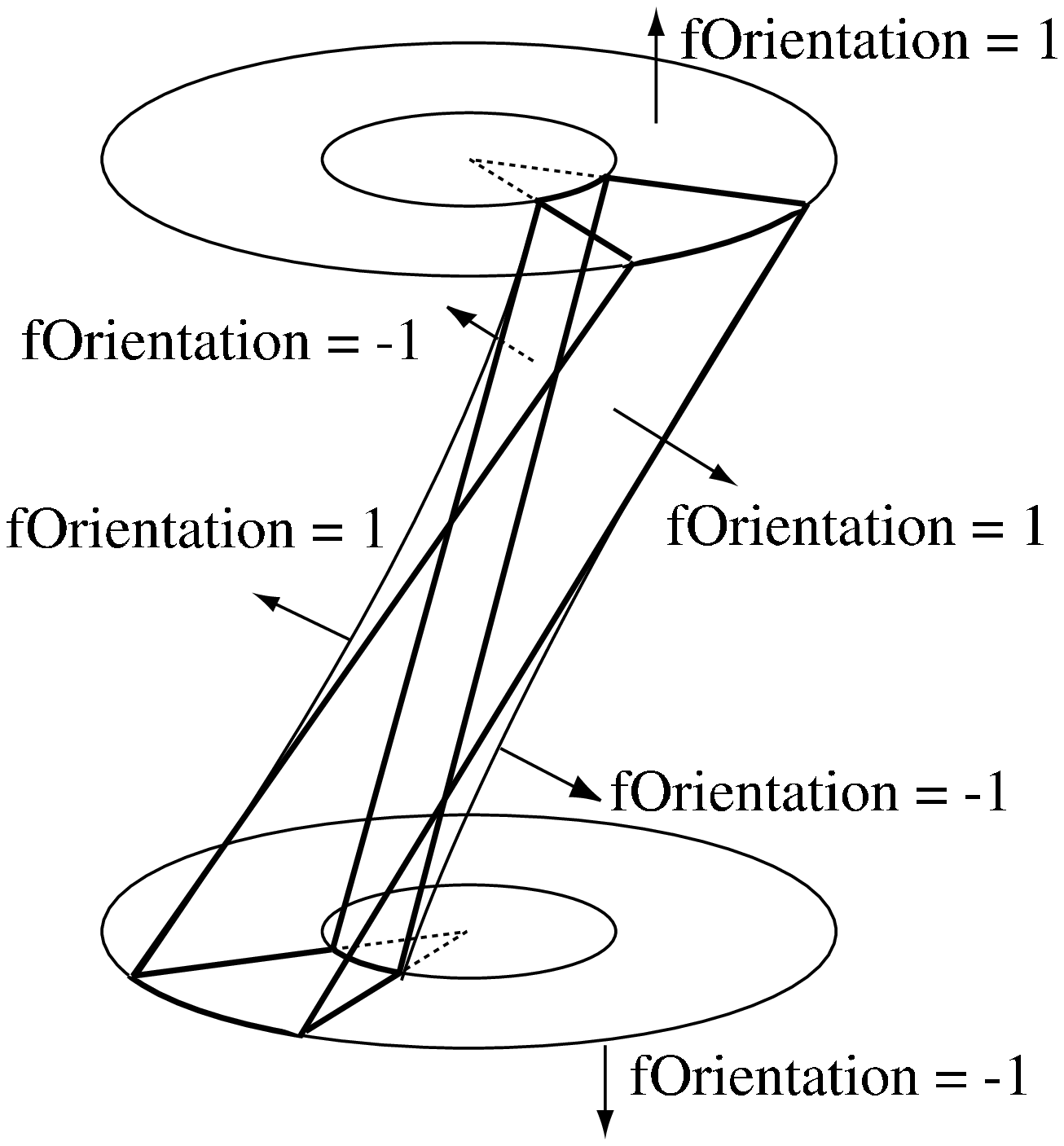}}
\caption[directry]{\label{Fig:Handedness}
Values of {\tt fOrientation} data member of {\tt J4TwistedTubs}
}
\end{minipage}
\end{center}
\end{figure}

Before coding our new solid "{\tt J4TwistedTubs}", we set the following guide line
to fulfill the required functionality discussed in the last section:
\begin{enumerate}
	\item {\tt J4TwistedTubs} should be implemented as a collection of bounding surfaces. It will thus have to have, as its data member, an array of pointers to the instances of the corresponding surface classes.
	\item All of these surface classes must inherit from an abstract class named "{\tt J4VSurface}", which carries generic information on each surface and sets basic interfaces to all the surface classes that inherit from it.
	\item Each surface is orientable and should know which side is outside, but the calculation of the distance to the surface from a point {\tt p} should not depend on whether the point is inside or outside of the solid, since the distance should only depend on the shape of the surface in question. 
	\item It is then the role of "{\tt DistanceToIn}" or "{\tt DistanceToOut}" methods of its parent {\tt J4VSurface} class to judge the distance depending on the context such as the angle between
the surface normal and the velocity vector {\tt v}. This way, we can avoid repetition of the same code.
	\item All {\tt J4TwistedTubs} has to do will then be to simply invoke "{\tt DistanceToIn}" or "{\tt DistanceToOut}" for each of the bounding surfaces and choose the best\footnote{
The best is usually the one that returned the shortest distance. One exception is the case in which a particle is coming into the solid from a corner or edge of the solid.
}. The surface normal can then be calculated by calling the corresponding {\tt GetNormal} method for the selected surface.
\end{enumerate}
As described in section \ref{Sec:geometry},
{\tt J4TwistedTubs} consists of three pairs of three kinds of surfaces.
We named these three kinds of bounding surfaces 
{\tt J4HyperboloidalSurface}, {\tt J4TwistedSurface}, and {\tt J4FlatSurface},
all of which are descendants of {\tt J4VSurface}. 
Interrelation of these classes is illustrated in Fig.~\ref{Fig:TwistedTubsUML}
as a class diagram.
On the other hand,  Fig.~\ref{Fig:Handedness} shows
the outward direction of each surface as stored in a data member
called {\tt fOrientation}.

As sketched above, the abstract base class {\tt J4VSurface} has the following methods: 
\begin{flushleft}
\underline{{\tt J4VSurface}}
\end{flushleft}
\begin{description}
	\item {\tt G4int DistanceToSurface(const G4ThreeVector \&p, G4ThreeVector *xx, \\
	G4double *distance, G4int *areacode) = 0 }~\\
		which calculates the distance ({\tt *distance}) from a point {\tt p} to the surface 
		as well as the point ({\tt *xx}) of the 
		closest approach to the surface and an {\tt *areacode} 
		given by {\tt GetAreaCode} explained below,
		and then returns the number of the points of closest approach  that is always 1.
		
	\item {\tt G4int DistanceToSurface(const G4ThreeVector \&p, const G4ThreeVector \&v, \\
	G4ThreeVector xx[], G4double distance[], G4int areacode[], \\
	G4bool isvalid[], EValidate validate) = 0}~\\
		which calculates the distance(s) ({\tt distance[]}) along the velocity vector {\tt v} 
		to the surface from the point {\tt p}, 
		the intersection(s) ({\tt xx[]}) of the particle track with the surface,
		and the areacode(s) ({\tt areacode[]})  of intersection(s), 
		and then returns the number of intersections.
		If so flagged by {\tt validate}, it does boundary check for each of the
		candidate intersection(s) and passes the test result through {\tt isvalid[]}.
		The return value then becomes the number of valid intersection(s).
	
	\item {\tt G4ThreeVector GetNormal(const G4ThreeVector \&xx, G4bool isglobal) = 0}~\\
		which calculates and returns the normal at a surface point {\tt xx}. 
		The surface point {\tt xx} and the normal to be returned are
		assumed to be given in the coordinate system of its mother solid
		or in the local frame of the surface in question, depending on
		whether {\tt isglobal} is {\tt TRUE} or not, respectively.  
		
	\item {\tt G4Int GetAreaCode(const G4ThreeVector \&xx, G4bool withTol) = 0}~\\
		which decides whether the intersection {\tt xx} 
		given by {\tt DistanceToSurface(p,v)}  
		is on a boundary, 
		or at a corner, or inside or outside of the surface
		with or without tolerance as flagged by {\tt withTol},
		and then returns a corresponding area code.
		
	\item {\tt G4double DistanceToIn(const G4ThreeVector \&p, const G4ThreeVector \&v, \\
				G4ThreeVector \&xx)}~\\
	         which invokes {\tt DistanceToSurface(p,v)} of its descendant concrete class 
		and  judges the validity of each of the resultant intersection(s) and the corresponding
		distance(s), 
		examining the sign of the distance, 
		the angle of the particle velocity {\tt v} to the surface normal 
		at the intersection point on the surface.  
		If the intersection is on a boundary or at a corner of the surface, 
		it also checks the angle to the surface normal(s) of the adjacent surface(s). 
		It then returns the distance if valid, or infinity otherwise,
		together with the valid intersection {\tt xx}, if any.
			 
	\item {\tt G4double DistanceToOut(const G4ThreeVector \&p, const G4ThreeVector \&v, \\
				G4ThreeVector \&xx)}~\\
	           which invokes {\tt DistanceToSurface(p)} of its descendant concrete class
		and judges the validity of each of the resultant intersections(s) and the corresponding
		distance(s), taking into account the sign of the distance 
		and the angle between the surface normal and the particle velocity {\tt v}, 
		and then returns the distance to the caller if valid,  or infinity otherwise,
		with the valid intersection passed through {\tt xx}, if any.
			   
	\item {\tt G4double DistanceTo(const G4ThreeVector \&p)}~\\
	         which just passes the return value of {\tt DistanceToSurface(p)} without any 
		additional judgment. 
		This functions is used in {\tt DistanceToIn(p)} or
		{\tt DistanceToOut(p)} of {\tt J4TwistedTubs}.
\end{description}

Notice that the first four of these member functions are pure virtual and
should be implemented in its derived concrete classes.
Since we have already explained how to construct a surface normal
in section \ref{Sec:geometry} and since description of {\tt GetAreaCode}
would only involve technical details,
we will concentrate, in what follows, on how we implemented the base class {\tt J4VSurface}
and the first two that are the {\tt DistanceToSurface} functions with 
and without the particle velocity,  for each of
the three derived surface classes: \\
{\tt J4HyperboloidalSurface}, {\tt J4TwistedSurface}, and {\tt J4FlatSurface}.

\section{Algorithm and Implementation}

\subsection{J4VSurface}

Being the base class for all the surface classes that make {\tt J4TwistedTubs}, 
the {\tt J4VSurface} class plays the following four major roles:
(1) it carries data members to store generic information on a surface, 
(2) it standardizes interface for generic methods to be implemented by its descendant,
(3) it provides tools commonly used by any derived surface class, and 
(4) it steers the distance calculation in {\tt DistanceToIn}
and {\tt DistanceToOut}, by filtering the return values from 
{\tt DistanceToSurface} of its derived class.
We now describe these points below in some more detail.

\subsubsection{Generic Data Members}
{\tt J4VSurface} has the following {\it protected} data members
to store generic information on its derived surface class.
\begin{itemize}
	\item Two axes of the surface
	\item Maximum and minimum limits of the surface along the two axes
	\item The rotation matrix and the translation vector to transform the surface
				from its mother solid ({\tt J4TwistedTubs}) coordinate system to the local frame attached to the surface
	\item The last value of the distance to the surface from a point {\tt p} 
	\item The last value of the surface normal
\end{itemize}
On the other hand, the next five data members are {\it private}, 
since they are basically used internally by {\tt J4VSurface}:
\begin{itemize}
	\item Four pointers to neighboring surfaces
	\item Four positions of the corner points
	\item An instance of a local class that stores information on the four boundary line segments
	\item The name of the surface
	\item A pointer to its mother solid ({\tt J4TwistedTubs})
\end{itemize}
Notice that  {\tt J4VSurface} currently assumes the existence of four corners and consequently four boundary line segments for its derived surface class.
It is, however, relatively easy to extend it to support variable number of
neighboring surfaces.

\subsubsection{Common Utility Functions}

\begin{description}
	\item {\tt G4double DistanceToLine(const G4ThreeVector \&p, const G4ThreeVector \&x0, \\
	const G4ThreeVector \&d, G4ThreeVector \&xx)}~\\
		to calculate the distance from a point {\tt p} to a line specified by 
		a reference point {\tt x0} on it and the line direction vector {\tt d},
		together with the point {\tt xx} of the closest approach on the line.
		It then returns the distance.
		
	\item {\tt G4double DistanceToPlane(const G4ThreeVector \&p, const G4ThreeVector \&x0, \\
	const G4ThreeVector \&n0, G4ThreeVector \&xx)}~\\
		to calculate the distance from {\tt p} to a plane specified by a normal {\tt n0}  
		and a reference point {\tt x0} on it,
		together with the point {\tt xx} of the closest approach to the surface.
		It then returns the distance.
		
	\item {\tt G4double DistanceToPlane(const G4ThreeVector \&p, const G4ThreeVector \&x0, \\
	const G4ThreeVector \&t1, const G4ThreeVector \&t2, G4ThreeVector \&xx, \\
	G4ThreeVector \&n)}~\\
		to calculate the distance from {\tt p} to a plane 
		and the point {\tt xx} of the closest approach to the plane, 
		where the plane is specified by a reference point {\tt x0} 
		and two linearly independent vectors {\tt t1} and {\tt t2} on the plane.
		It then returns the distance.
		
	\item {\tt G4double DistanceToBoundary(G4int areacode, G4ThreeVector \&xx, \\
	const G4ThreeVector \&p)}~\\
		to calculate the distance from {\tt p} to a boundary 
		specified by {\tt areacode}, and the point {\tt xx} of the closest approach
		to the boundary.
		Its return value is the distance. 
		
	\item {\tt G4int AmIOnLeftSide(const G4ThreeVector \&me, const G4ThreeVector \&ref, \\
	G4bool withTol)}~\\
		to judge relation between two vectors {\tt me} and {\tt ref} 
		by comparing their $\phi$ angles. 
		If {\tt me} lies clearly on the left side of {\tt ref}, namely {\tt me} 
		has a $\phi$ angle smaller than that of {\tt ref}, it returns 1. 
		If {\tt withTol} is {\tt TRUE} and if {\tt me} lies 
		within $\pm 0.5 \times${\tt kAngTolerance} ($=0.5 \times 10^{-9}$) radians of {\tt ref}, 
		it returns 0. Otherwise, it returns -1.
\end{description}

\subsubsection{DistanceToIn and  DistanceToOut}

\noindent
\underline{{\tt DistanceToIn(p,v)}}
\vspace*{0.5cm}

\begin{figure}
\begin{center}
\begin{minipage}{5cm}
\centerline{
\epsfxsize=4.7cm
\epsfbox{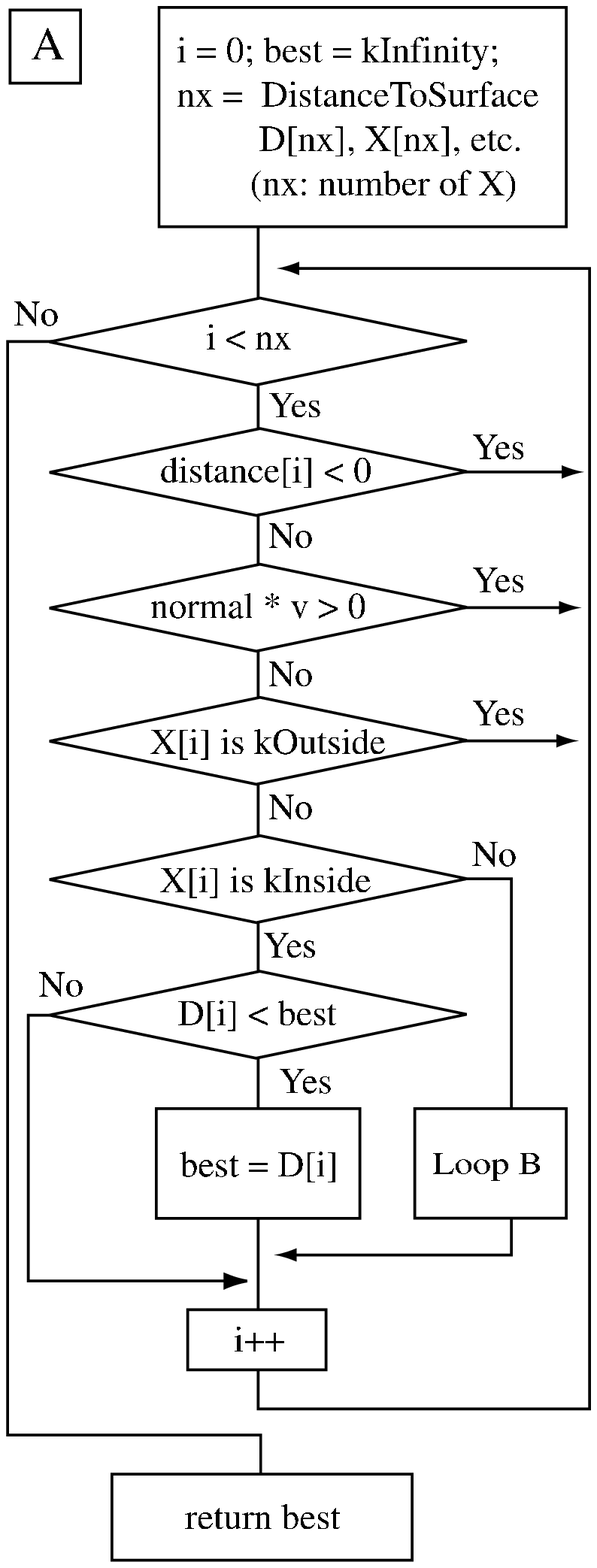}}
\end{minipage}
\vspace*{0.5cm}
\begin{minipage}{5cm}
\centerline{
\epsfxsize=4.7cm
\epsfbox{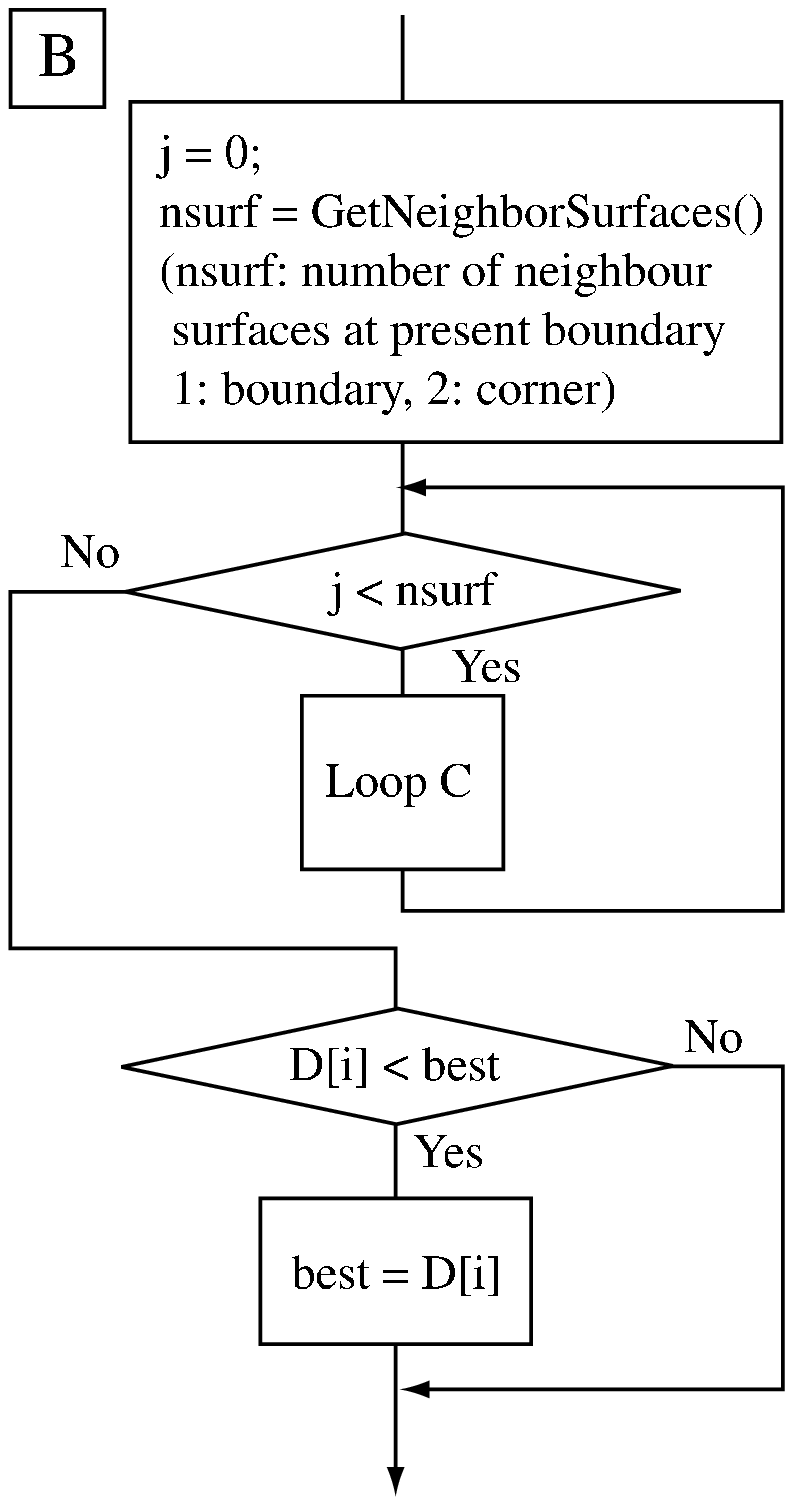}}
\end{minipage}
\vspace*{0.5cm}
\begin{minipage}{5cm}
\centerline{
\epsfxsize=4.7cm
\epsfbox{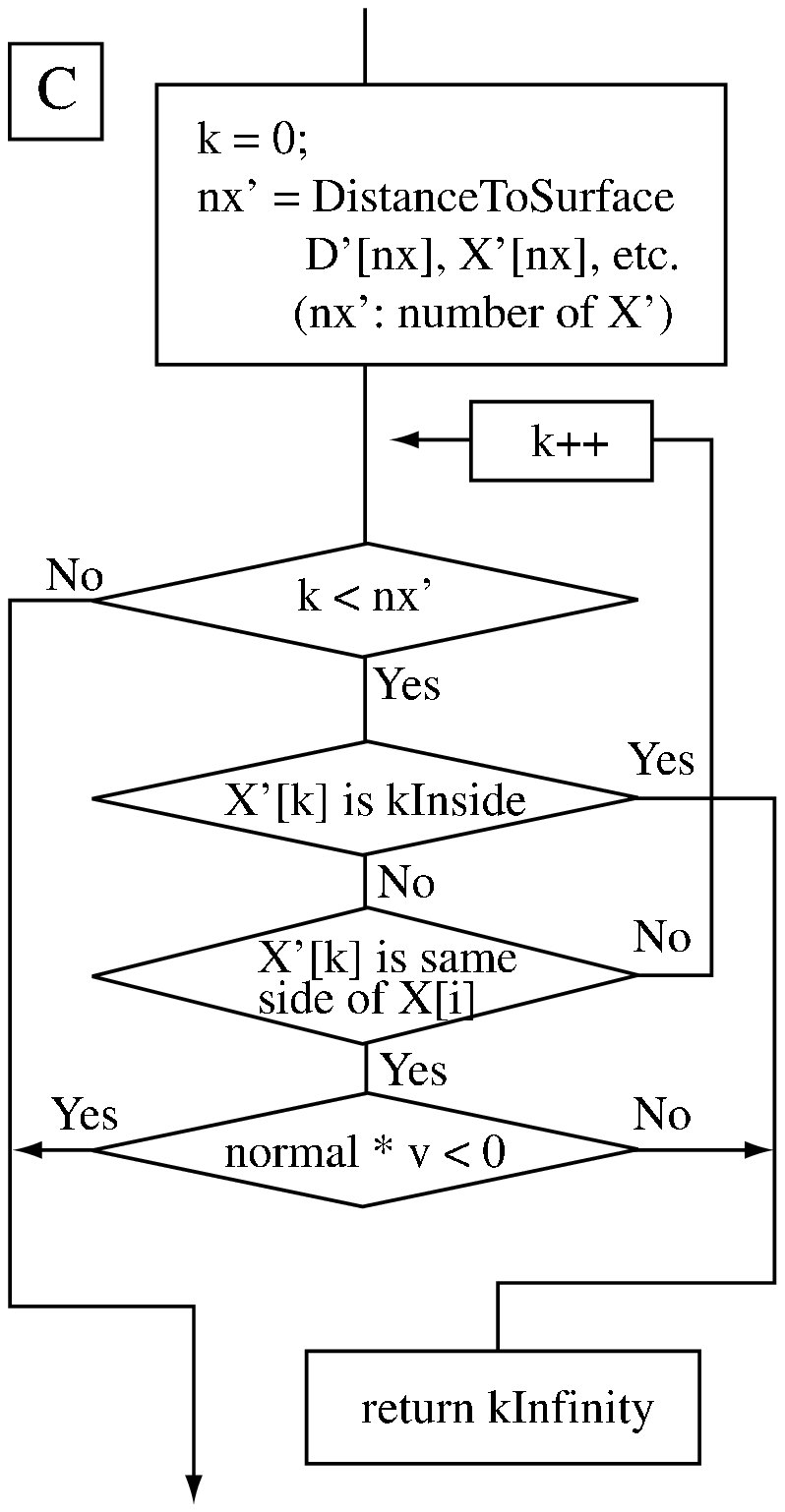}}
\end{minipage}
\caption[directry]{\label{Fig:flowchart}
Flow chart of DistanceToIn
}
\end{center}
\end{figure}

On invocation, {\tt DistanceToIn(p,v)}\footnote{
In what follows, we often leave out arguments which are irrelevant
in the explanation of algorithmic aspects of each function,
as long as there is no possibility of confusion.
}
proceeds according to the following algorithm: 
\begin{enumerate}
	\item Initialize a temporary data buffer to store the current best distance and intersection to {\tt kInfinity}.
	\item Then call {\tt DistanceToSurface(p,v)} and get a set of distance(s) ($D$) and intersection(s) ($\mathbold{X}$): 
				the number of intersections is at most 2 in our case.
	\item If a candidate $D$ and its corresponding $\mathbold{X}$ satisfy the following conditions,  
				discard this candidate and move on to the next, if any:
		\begin {itemize}
			\item $D$ is negative, or
			\item the scalar product between the outward surface normal 
				at $\mathbold{X}$ and the particle velocity {\tt v} is positive, or
			\item $\mathbold{X}$ is outside of the surface.
		\end {itemize}
	\item If $\mathbold{X}$ is inside of the surface and if $D$ is smaller than the last value stored 
				in the temporary data buffer defined in step 1, update the stored $D$ and $\mathbold{X}$ 
				and move on to the next intersection, if any.
	\item If $\mathbold{X}$ is on the boundary of the surface within $\pm 0.5 \times${\tt kCarTolerance}($=0.5 \times 10^{-9}$), 
				we need to judge whether the particle is really entering the solid or just grazing the solid and flying away. 
				In this case, invoke the {\tt DistanceToSurface(p,v)}  method of  the (two) neighboring surface(s) 
				(two if the intersection is at a corner, one otherwise) and get a set of distance(s) ($D'$) and intersection(s) ($\mathbold{X}'$), 
				and if $\mathbold{X}'$ satisfies the following conditions, exit {\tt DistanceToIn}  immediately by returning {\tt kInfinity}
				as the distance to the surface originally in question.
		\begin{itemize}
			\item If $\mathbold{X}'$ is inside of a neighboring surface\footnote{
				In this case this neighboring surface should be the one through which the particle enters the solid.
			}, or
			\item if the $\mathbold{X}'$ is on the same boundary (or at the same corner) as 
			$\mathbold{X}$ and if the scalar product of the outward surface normal at 
			$\mathbold{X}'$ and the particle velocity {\tt v} is positive\footnote{
				A particle cannot come into a solid from the edge (or corner) if the scalar product is positive 
				for any of the surfaces forming the edge (or corner).
			}.
		\end{itemize}
	\item If $\mathbold{X}$ 
	survives  all the tests for all the neighboring surfaces, 
	check whether the current $D$ is smaller than that stored in the temporary data buffer,
	and if so, update the stored $D$ and $\mathbold{X}$.
	\item Return the stored distance in the temporary data buffer.
\end{enumerate}
The flow chart of {\tt DistanceToIn} is shown in Fig.~\ref{Fig:flowchart}.

\vspace*{0.5cm}

\noindent
\underline{{\tt DistanceToOut(p,v)}}

\vspace*{0.5cm}
On the other hand, the algorithm for {\tt DistanceToOut(p,v)} is less complicated as
described below:
\begin{enumerate}
	\item Initialize a temporary data buffer to store the current best distance and  intersection to {\tt kInfinity}.
	\item Call {\tt DistanceToSurface(p,v)} and get a set of distance(s) ($D$) and intersection(s) ($\mathbold{X}$): the number of intersections is at most 2 in our case.
	\item If $D$ and $\mathbold{X}$ meet the following conditions, reject this candidate and move on to the next:
		\begin {itemize}
			\item $D$ is negative, or
			\item the scalar product of the outward surface normal at $\mathbold{X}$ and the particle velocity {\tt v} is negative, or
			\item $\mathbold{X}$ is outside of the surface.
		\end {itemize}
	\item If $D$ is 
				smaller than the stored value in the temporary buffer, update the stored $D$ and $\mathbold{X}$. 
	\item Return the best distance stored in the temporary buffer.
\end{enumerate}

\vspace*{0.5cm}
\noindent
\underline{{\tt DistanceTo(p)}}\\

\noindent
{\tt DistanceTo(p)}, which is called by the {\tt DistanceToIn(p)} and {\tt DistaceToOut(p)} methods of {\tt J4TwistedTubs}, 
just returns {\tt DistanceToSurface(p)} which is implemented in the derived surface class.

\subsection{J4HyperboloidalSurface}

Geant4 provides a solid class called  {\tt G4Hype}
to represent a hyperboloidal volume 
with its symmetry axis parallel to the $z$-axis.
Since {\tt J4HyperboloidalSurface} follows the basic algorithm 
to calculate the inward or outward normal and the minimal
distance to a hyperboloidal surface,
we only sketch it here.

\subsubsection{DistanceToSurface(p, v)}

The distance to a hyperboloidal surface along a velocity 
${\tt v} = (v_x, v_y, v_z)$ from a point ${\tt p} = (p_{x}, p_{y}, p_{z})$
can be obtained as follows.
Define the intersection of a particle track
with the hyperboloidal surface to be 
$\mathbold{X} = {\tt p} + t {\tt v} =  (p_{x} + t v_x, p_{y} + t v_y, p_{z} + t v_z)$,
where $t$, the time, equals the distance we want, provided that
${\tt v}$ is normalized to unity.
From Eq.~\ref{Eq:phi(z)}, we then have a quadratic equation for $t$:
\begin{equation}
(p_{x} + t v_x)^2 + (p_{y} + t v_y)^2 = r_{z=0}^2 + (p_{z} + v_z)^2 \tan^2 \alpha.
\label{Eq:disttohype}
\end{equation}

\subsubsection{DistanceToSurface(p)}

{\tt DistanceToSurface(p)} is to calculate the minimum distance to 
the surface in question.
Since this function is primarily used to set the safety radius
not to cross any volume boundaries,
it is allowed to underestimate the minimum distance.
In the calculation of this minimum distance we can thus
approximate the surface curve by a properly chosen
straight line.
Taking advantage of $z$-symmetry, we only have to 
consider the region of $z \ge 0$.

\begin{flushleft}
\underline{Distance from Outside}
\end{flushleft}

\begin{figure}
\begin{center}
\begin{minipage}{6cm}
\centerline{
\epsfysize=6cm
\epsfbox{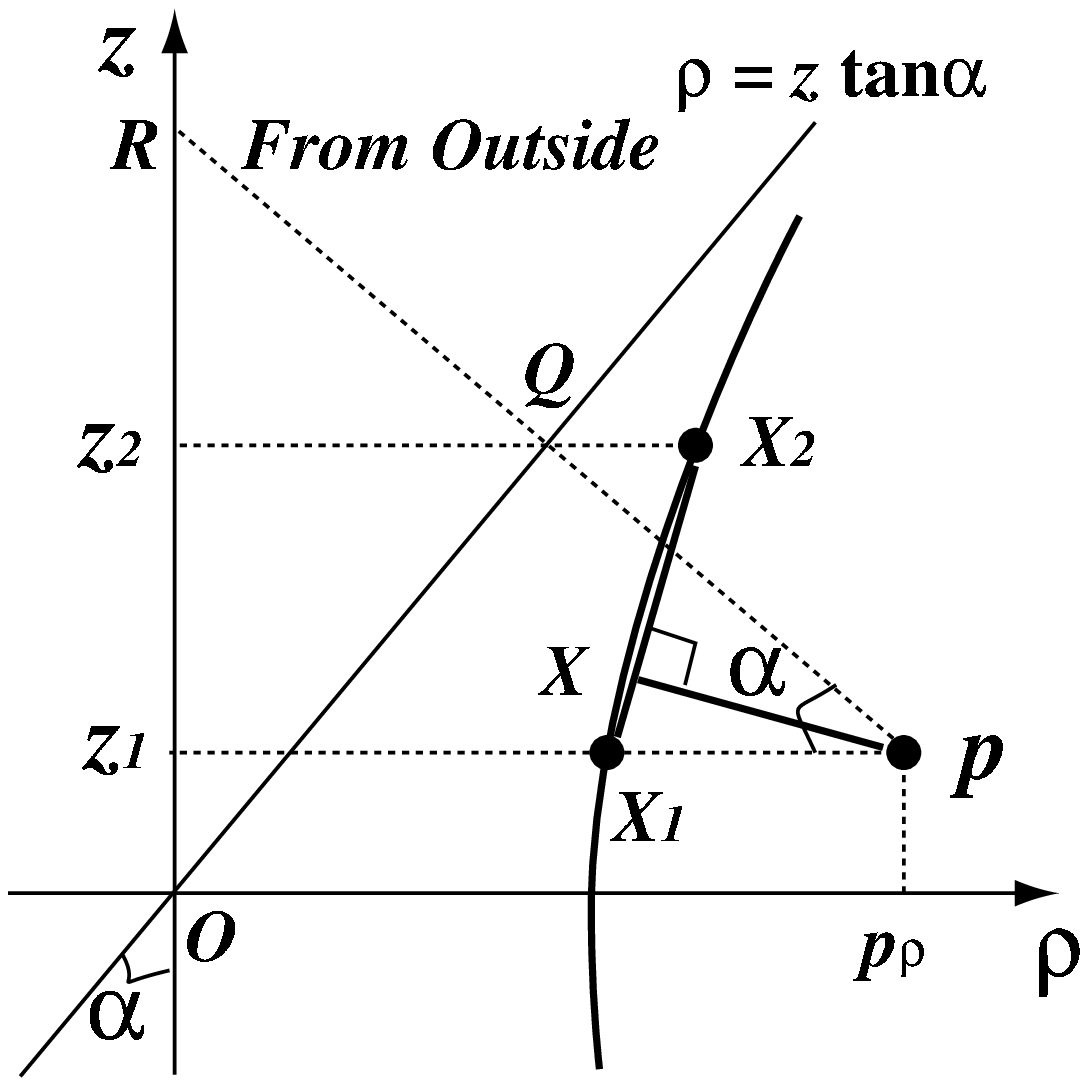}}
\caption[directry]{\label{Fig:HypenormalOutward}
Distance from outside
}
\end{minipage}
\hspace*{1cm}
\begin{minipage}{6cm}
\centerline{
\epsfysize=6cm
\epsfbox{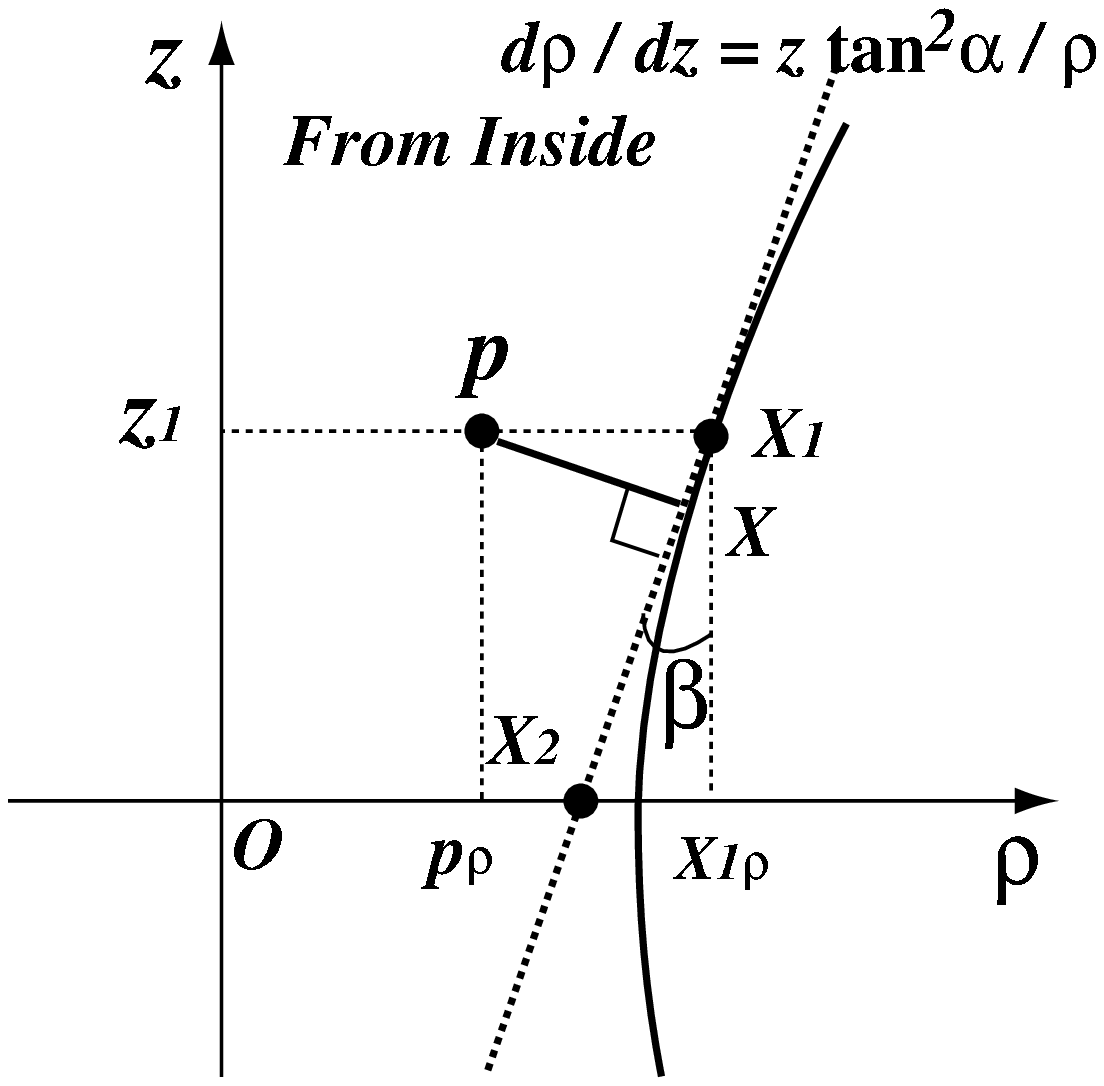}}
\caption[directry]{\label{Fig:HypenormalInward}
distance from inside
}
\end{minipage}
\end{center}
\end{figure}

\noindent
In this case, the proper line can be chosen in the following way:
\begin{enumerate}
	\item Find the surface point ($\mathbold{X}_1$) which has the same $z$ as {\tt p}.
	\item Draw a line from {\tt p} which is perpendicular to the asymptotic line $\rho = z~\tan \alpha$ and let their intersection be $Q$.
	\item Find the surface point ($\mathbold{X}_2$) that has the same $z$ as $Q$.
	\item Connect $\mathbold{X}_1$ and $\mathbold{X}_2$
\end{enumerate}
Then, by construction, the distance from {\tt p}  to the line $\mathbold{X}_1 \mathbold{X}_2$
never exceeds the minimum distance to the hyperboloidal surface.
The $z$ component of $\mathbold{X}_2$ can be obtained geometrically 
from Fig.~\ref{Fig:HypenormalOutward}:
\begin{eqnarray}
z_2 &=& OQ~\cos\alpha \cr
OQ &=& OR~\cos\alpha \cr
OR &=& z_1 + p_\rho~\tan\alpha.
\end{eqnarray}

\begin{flushleft}
\underline{Distance from Inside}
\end{flushleft}

\noindent
In this case, a proper line will be the tangential line
at the surface point $\mathbold{X}_1$ which has the same $z$ component
as {\tt p} (see Fig.~\ref{Fig:HypenormalInward}). 
The tangential vector is given by  Eq.~\ref{Eq:HypeTangential}.

\subsection{J4FlatSurface}

\subsubsection{DistanceToSurface(p, v)}

The distance to a plane along a velocity vector {\tt v} 
from a point {\tt p} is readily obtained from 
\begin{equation}
t = - \mathbold{n} \cdot (\mathbold{p} - \mathbold{x}_0) / |\mathbold{n}|,
\label{Eq:disttoflat}
\end{equation}
where \mathbold{n} is the normal to the plane and $\mathbold{x}_0$
is the reference point on the plane.

\subsubsection{DistanceToSurface(p)}

Since the two planar boundary surfaces of {\tt J4TwistedTubs}
are perpendicular to the $z$-axis,
which means its normal is in the direction of the $z$-axis.
All {\tt DistanceToSurface(p)} has to do is simply to return the
difference between $z$-components of {\tt p} and the plane in question.

\subsection{J4TwistedSurface}

\subsubsection{DistanceToSurface(p, v)}

The distance to a twisted surface from
a point ${\tt p} = (p_x, p_y, p_z)$  
along a velocity vector ${\tt v} = (v_x, v_y, v_z)$ can be obtained 
in a similar manner to that for {\tt J4HyperboloidSurface}.
Substituting $\mathbold{X} = (X_x, X_y, X_z)= {\tt p} + t {\tt v}$ in
Eq.~\ref{Eq:Twisted}, we have the following quadratic equation:
\begin{equation}
(\kappa v_x v_z) t^2 + ((v_x p_{z} + v_z p_{x}) \kappa - v_y) t + \kappa p_{x} p_{z} - p_{y}  = 0
\label{Eq:twistedsurface}
\end{equation}
for the distance $t$ to the surface.

Notice that, although this quadratic equation may contain a solution 
for which $X_x = \rho_c < 0$,
such a situation should not take place in practice,
since the twist angle cannot exceed $\Delta \phi = \pi$
without breaking wires.
This requires $X_x = \rho_c$ in Eq.~\ref{Eq:Twisted} be positive.
Notice also that, for a given stereo angle,
the twist angle $\Delta \phi$ increases with the wire length.
In solving Eq.~\ref{Eq:twistedsurface}, we thus need to set
limits on $z$ as determined by the endcap locations.
These validity checks can be made
for the obtained crossing point $\mathbold{X}$ 
after solving the above equation.

\subsubsection{DistanceToSurface(p)}

\begin{figure}
\begin{center}
\centerline{
\epsfxsize=15cm
\epsfbox{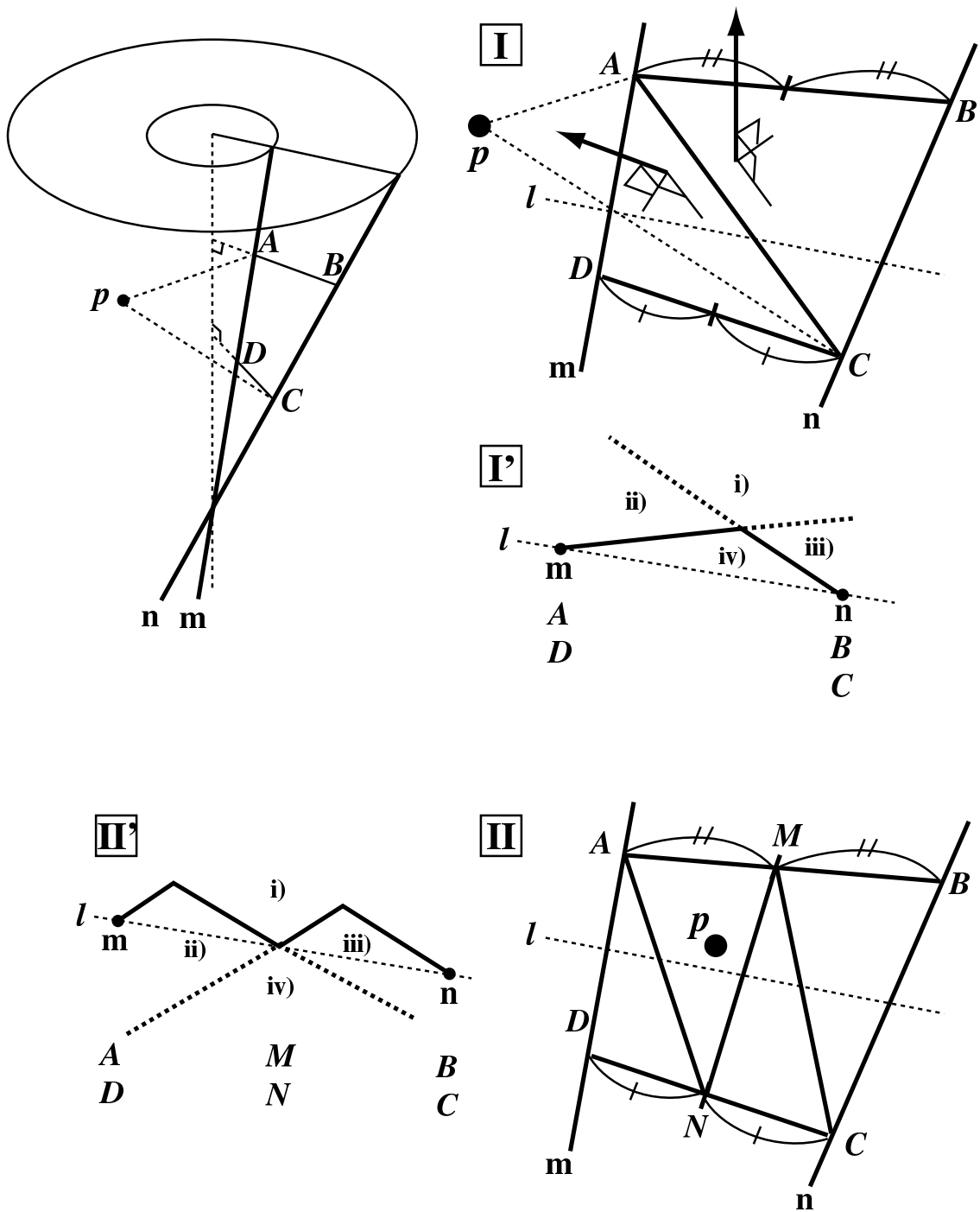}}
\caption[directry]{\label{Fig:TwistedSurface2}
Procedure to choose an appropriate plane to approximate 
the minimum distance to the twisted surface
}
\end{center}
\end{figure}

The exact equation for the minimal distance to the twisted surface
from a point {\tt p} is biquadratic.
As in the case of {\tt J4HyperboloidalSurface},
what we really need is, however, not the exact minimal distance.
Its approximation will do, provided that it never exceeds the true
minimal distance.  
We thus developed an approximation method that uses the distance 
to an appropriately chosen plane instead of solving the exact equation.
Now the question is how we find such a plane to
approximate the twisted surface.
Fig.~\ref{Fig:TwistedSurface2} illustrates the procedure to find it.

Since we should never overestimate the distance,
the plane has to lie on the same side as the point {\tt p}
with respect to the twisted surface, at least in the vicinity of the point of the
closest approach to the plane.
In order to fulfill this condition, essential is the following observation
about the geometric features of the twisted surface.
As stressed in section \ref{Sec:geometry} 
the twisted surface contains two kinds of straight lines
given by ${\bf e}_z$ and ${\bf e}_x$ in Eq.~\ref{Eq:TwistedNormal}:
one that can be regarded as one of 
the field wires making up the twisted surface
and the other that can be taken as a $\phi$ boundary of a
$z$-slice of a stereo cell.
Lines $AD$ and $BC$ of
Fig.~\ref{Fig:TwistedSurface2} are the first type (${\bf e}_z$-type),
while $AB$ and $DC$ are the second type (${\bf e}_x$-type).
We can thus cut out, from the twisted surface, 
a segment $ABCD$ that is bounded by four straight lines.
Recall that this segment has a saddle shape:
it is concave along the diagonal $AC$ 
in Fig.~\ref{Fig:TwistedSurface2},
while it is convex along the other diagonal $BD$\footnote{
Of course, which is convex or concave flips, depending on
from which side you are looking.}.

Our goal is to choose a plane that is placed in front of both 
the current point {\tt p} and the twisted surface
and to calculate the distance to that plane.
Once we find a diagonal which does not cross the twisted surface, 
we can immediately span a plane with the diagonal and 
an adjacent side ($AB$ or $DC$) which is entirely lies on the surface\footnote{
Remember the twisted surface can also be regarded as a locus of line $AB$ in Fig.~\ref{Fig:TwistedSurface}.
}.
Hence our first task is to select an appropriate diagonal along which
the surface is concave.  
The procedure in {\tt DistanceToSurface(p)} to select such a diagonal can be
itemized as follows:
\begin{enumerate}
	\item Let $A$ and $C$ be the points of the closest approach from the point {\tt p}
		to the inner and outer boundary lines 
		($m$ and $n$ in Fig.~\ref{Fig:TwistedSurface2}-I).
		If the $z$ components of $A$ and $C$ do not sandwich 
		the $z$ component of {\tt p},
		move $A$ or $C$ along the boundary lines until its $z$
		component matches that of {\tt p}.		
	\item Choose $D$ and $B$ on lines $m$ and $n$, respectively,
		in such a way that lines $AB$ and $DC$ are perpendicular to the
		$z$-axis and thus entirely contained in the twisted surface.
	\item First check if diagonal $AC$ satisfies our requirement
		by examining the sign of product of the following three test variables:
		\begin{enumerate}
			\item the signed twist angle $\Delta \phi$,
			\item an orientation index, which is 1
				if {\tt p} is on the negative $\phi$ side of the  twisted surface,
				0 if {\tt p} is on the surface within $\pm 0.5 \times${\tt kAngTolerance} 
				($=0.5 \times 10^{-9}$) radians, and
				-1 otherwise, and
			\item the $z$ coordinate of $A$ minus that of $C$. 
		\end{enumerate}
		If the product is positive, accept the diagonal.
		If it is 0, which means either (b) is 0 or (c) is 0.
		If (b) is 0, {\tt p} is on the surface (or is very close to the surface).
		In this case, immediately exit {\tt DistanceToSurface(p)} by returning 0
		as the minimal distance to the twisted surface.
		Else if (c) is 0, return the distance to line $AC$ from {\tt p}. 
		Finally if the product turns out to be negative, 
		which means that the diagonal is invalid crossing the twisted surface,
		exchange $A$ and $D$, and $C$ and $B$, respectively.
\end {enumerate}
Diagonal $AC$ can now be used to set up two planes:  plane $ADC$ and $ABC$.
The normal to these planes are indicated in Fig.~\ref{Fig:TwistedSurface2}-I,
and should point into the hemisphere that contains {\tt p}.
The subsequent distance
calculation has two branches, depending on the relative location of {\tt p}
with respect to these two candidate planes.

\vspace*{0.5cm}
\noindent
\underline{Case A:  {\tt p} is in region i), or ii), or iii)  of Fig.~\ref{Fig:TwistedSurface2}-I'}\\

\noindent
In these cases, all {\tt DistanceToSurface(p)} needs to do is to return
the smallest positive distance.

\vspace*{0.5cm}
\noindent
\underline{Case B: {\tt p} is in region vi)  of Fig.~\ref{Fig:TwistedSurface2}-I'}\\

\noindent
In this case, we must re-define the planes so that at least one of them lies 
behind point {\tt p}.
Let $M$ and $N$  be the middle points of $AB$ and $CD$,
respectively.
Since line $MN$ is fully contained in the twisted surface, 
we can now define new quadrangles $ADNM$ and $MNCB$.
We can then define four planes $ADN$, $AMN$, $MNC$, and $MBC$.
Planes $ADN$ and $MBC$, however, coincide with
$ADC$ and $ABC$, respectively, and have hence been tested already.
Therefore, we just have to test the remaining two planes.
Again there are two cases to consider, depending on the relative
location of point {\tt p} with respect to the planes
(see Fig.~{\ref{Fig:TwistedSurface2}}-I'),

\begin{description}
\item[{\rm Case B-1: {\tt p} is in region i) of Fig.~\ref{Fig:TwistedSurface2}-II'}]~\\
	In this case, {\tt DistanceToSurface(p)} should  just return the smaller of the  distance to 	$AMN$ and that to $MNC$.
\item[{\rm Case B-2: p is in region ii) or iii)  of Fig.~\ref{Fig:TwistedSurface2}-II'}]~\\
	In these cases, the situation is essentially the same as in iv) of 
	Fig.~\ref{Fig:TwistedSurface2}-I'.
	Compare the distances to $AMN$ and $MNC$ and split into two
	the one that gave a negative distance.
\end{description}
Notice that, by construction, {\tt p} must not fall into region iv)
of Fig.~\ref{Fig:TwistedSurface2}-II'.
If it happens, therefore, we just abort the program.

\noindent
The procedure for B-2 is actually implemented as a recursive call:
{\tt J4TwistedSurface} is equipped with 
its own {\tt DistanceToPlane}, which
calls itself recursively until the situation changes to B-1
and returns a positive distance.
In this way {\tt DistanceToSurface} can get the desired distance by
a single call to {\tt DistanceToPlane},
when case B takes place.

\section{Test of J4TwistedTubs}

\begin{figure}
\begin{center}
\centerline{
\epsfxsize=6cm
\epsfbox{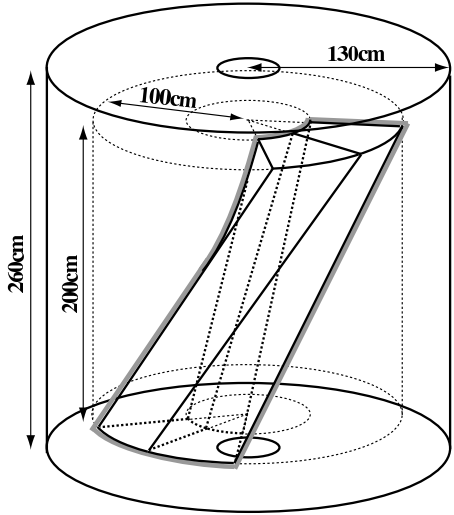}}
\caption[directory]{\label{Fig:testgeom}
Schematic view of the test geometry
}
\end{center}
\end{figure}

In order to test the new solid class {\tt J4TwistedTubs}, 
we prepared a test program (for stereo cells) as follows.
We first constructed a world volume with a cubic shape of 6m $\times$ 6m $\times$ 6m
and put, at its center, a cylindrical tube-type layer with inner and outer radii of 30cm
and 130cm, respectively, and a length of 260cm.
In this cylindrical layer,
we placed a {\tt J4TwistedTubs} object, which has
a twist angle ($\Delta \phi$) and a $\phi$-width of $\pi / 3$, inner and outer radii
at the endcaps of 50cm and 100cm, respectively, and a $z$-length of 200cm.
Into this mother {\tt J4TwistedTubs} volume,
installed were two daughter {\tt J4TwistedTubs} objects 
which have half the $\phi$-width but otherwise the same geometry.
All of these volumes are made of air.
A schematic view of the test geometry is shown in Fig.~\ref{Fig:testgeom}.

In order to evaluate the computing speed of {\tt J4TwistedTubs},
we also prepared another test program (for axial cells) with 
the mother and the two daughter {\tt J4TwistedTubs}
objects replaced by ordinary cylindrical tubs ({\tt G4Tubs}).

\subsection{Test with Geantinos}

\begin{figure}
\begin{center}
\centerline{
\epsfxsize=15cm
\epsfbox{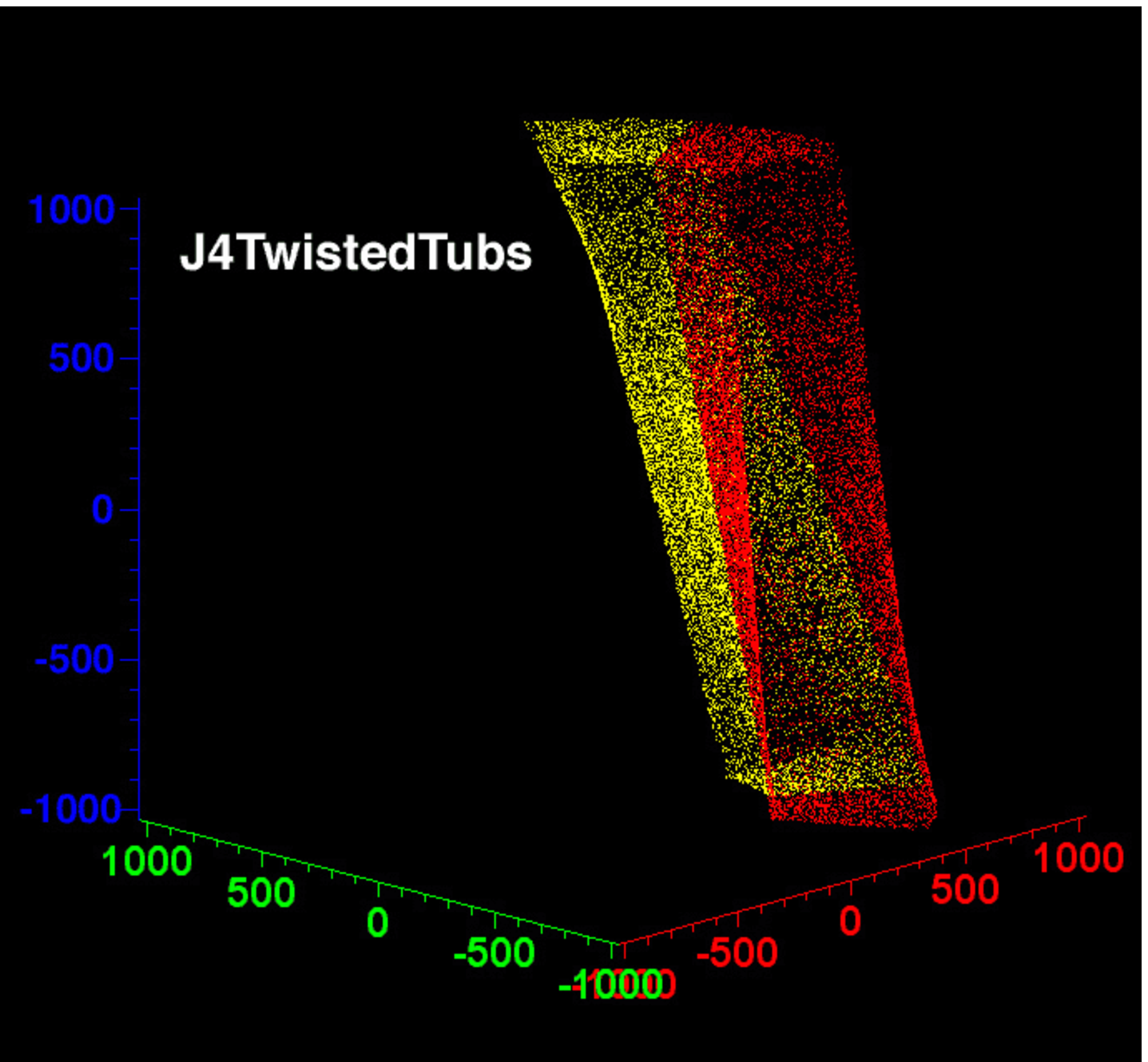}}
\caption[directory]{\label{Fig:thebe}
3-dimensional view of hit points of geantinos 
on the daughter {\tt J4TwistedTubs} volumes
}
\end{center}
\begin{center}
\begin{minipage}{7.5cm}
\centerline{
\epsfxsize=7.5cm
\epsfbox{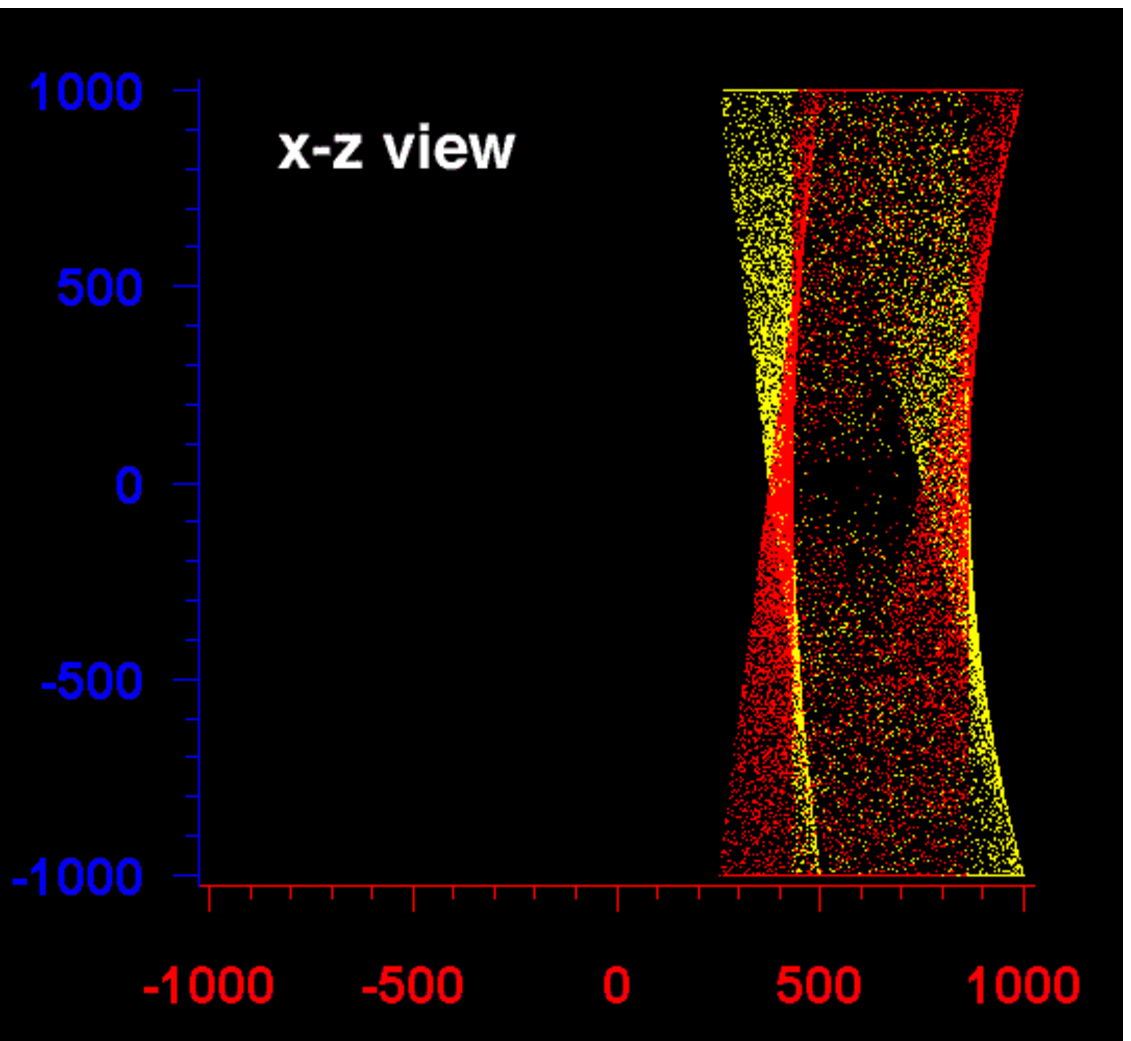}}
\caption[directory]{\label{Fig:thebex-z}
$x$-$z$ view
}
\end{minipage}
\hspace*{0.3cm}
\begin{minipage}{7.5cm}
\centerline{
\epsfxsize=7.5cm
\epsfbox{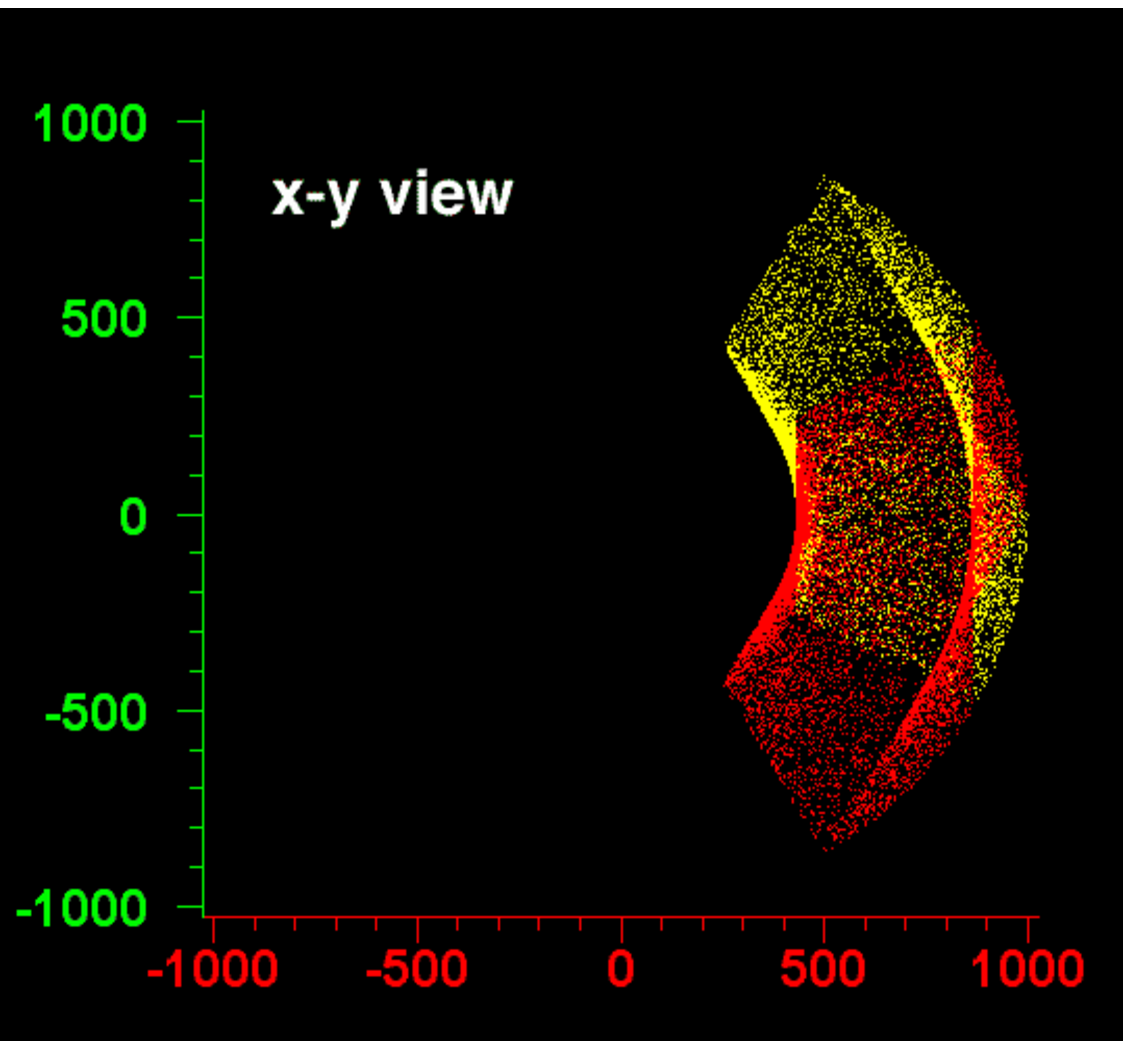}}
\caption[directry]{\label{Fig:thebex-y}
$x$-$y$ view
}
\end{minipage}
\end{center}
\end{figure}

Geant4 provides a hypothetical particle called "{\it geantino}"
designed primarily for geometry tests.
A geantino does not  interact with any materials and flies
along a straight line trajectory.
It only makes a hit when it crosses a boundary of volumes
which are {\it sensitive}\footnote{Geant4 generates a hit when (1) a physics process 
took place on a flying particle or (2) its track crossed
a boundary of a volume registered as a sensitive detector.}.
We first carried out a geantino test of {\tt J4TwistedTubs},
using the aforementioned geometry test program
with only the daughter {\tt J4TwistedTubs} objects made
sensitive.
Fig.~\ref{Fig:thebe} is a 3-dimensional view of 
hit points made by 10000 geantinos  injected from
$4\pi$ steradian around the surrounding cylinder.
As seen from the figure all the hit points 
are correctly on the surface of 
the daughter {\tt J4TwistedTubs} volumes.

\subsection{Test with Higgs Events}

Unlike geantinos, a real charged particle makes a curved trajectory
in a central tracker due to the magnetic field applied to it,
and interacts with various detector materials,
producing, for instance, low energy tracks such as 
$\delta$-rays which might complicate
the particle tracking through the detector volumes.
In order to stress-test our {\tt J4TwistedTubs} 
under more realistic environment,
we also tried Higgs events ($e^+e^- \rightarrow ZH$ at $\sqrt{s} = 350~{\rm GeV}$)
generated by Pythia\cite{Ref:pythia}.
In this case, the number of tracks with a transverse momentum of 1~GeV or greater
is around 50 per event.
Because of the lack of a calorimeter in this test program,
however, relatively low energy tracks curl up and pass through the sensitive volumes
multiple times.
Taking into account this curl-up effect and
the geometrical acceptance of the {\tt J4TwistedTubs} volumes being about $(\pi/3)/(2\pi) = 1/6$,
the average number of tracks passing through the twisted volumes 
was estimated to be around 10 per event.
Processing of 1000 Higgs events,
which thus correspond to roughly 10000 tracks hitting the sensitive volumes,
took 11647 seconds 
on a Power Macintosh 800MHz (Dual CPU) with 1GB memory
for the {\tt J4TwistedTubs} test program.
The same test took 9242 seconds on the same platform for the axial cell test program.
The 25\% extra CPU time consumed by {\tt J4TwistedTubs} is 
acceptable for our purpose,
considering the expected CPU time necessary for calorimeter simulation.

\section{Discussion}

So far we treated the twisted surface analytically
in {\tt J4TwistedSurface}.
We found, however, Eq.~\ref{Eq:twistedsurface},
which is the key equation for our analytic treatment, 
suffers from roundoff errors in the situation sketched below. 

In general a roundoff error enters as a result of a subtraction 
of two numbers that share many common digits.
In the case of {\tt J4TwistedSurface}, this happens when
 ${\tt v}_x$ or ${\tt v}_z$ or both become small,
and consequently the second term of the discriminant 
of the quadratic equation Eq.~\ref{Eq:twistedsurface}
becomes negligible.
Geometrically ${\tt v}_x = 0$ corresponds to an
extreme case in which the velocity vector of the particle
is contained in a plane that is spanned by
a straight line in the direction of a wire
and another straight line which is parallel with the $z$-axis.
On the other hand, ${\tt v}_z = 0$ is the case 
in which the velocity vector is in the plane 
that is perpendicular to the $z$-axis.
In both of these cases, the problem becomes purely 2-dimensional
and hence the equation becomes linear.

When the quadratic equation is used in such a case, 
{\tt DistanceToSurface(p, v)} might 
return a crossing point that falls short of or goes beyond the surface
by more than {\tt kCarTolerance}.
The next step might then end up with an impossible
situation where {\tt DistanceToIn} ({\tt DistanceToOut}) 
would be called from inside (outside) of the volume, 
confusing the tracking program in Geant4.

\begin{figure}
\begin{center}
\centerline{
\epsfysize=8cm
\epsfbox{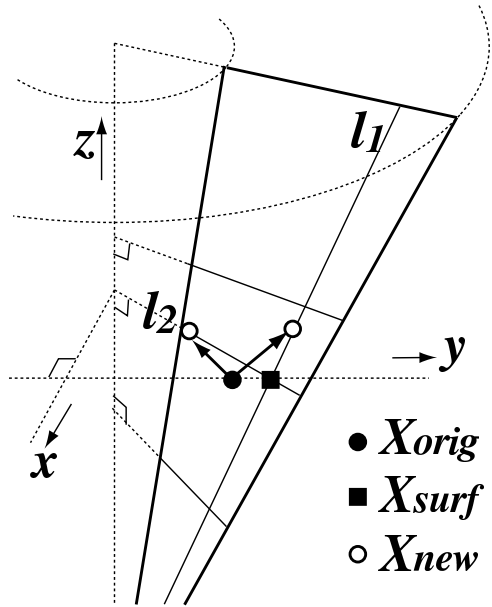}}
\caption[directry]{\label{Fig:discussion}
Procedure to fix the estimated intersection ($X_{orig}$)
when it falls short of the surface due to a roundoff error.
The surface point ($X_{surf}$) has the same $x$ and
$z$ coordinates as $X_{orig}$.
We approximate the twisted surface by the tangential plane
at $X_{surf}$, which is spanned by
two straight line sections $l_1$ and $l_2$,
and calculate a new crossing point
($X_{new}$) with this plane.
In the limit of $v_z = 0$ or $v_x = 0$, for which the numerical
instability is expected, the new intersection will be on
$l_2$ or $l_1$, respectively, and therefore the approximation
will become exact.
}
\end{center}
\end{figure}

When the calculated crossing point ($X_{orig}$ in Fig.~\ref{Fig:discussion}) is found 
to be more than {\tt kCarTolerance}-off the surface,
we apply the following Newtonian method
in order to avoid such inconsistency. 
We first approximate the twisted surface, at the surface point
that has the same local $x$ and $z$ coordinates ($X_{surf}$), 
by a tangential plane
spanned by two straight lines contained in the twisted surface: 
one in the wire direction and the other in the radial direction.
We then recalculate the crossing point ($X_{new}$) with this tangential plane.
We iterate this procedure until the recalculated crossing point
is found on the surface within the tolerance.

\section{Summary and Conclusion}
\label{Sec:conclusions}
We have developed a new Geant4 solid called {\tt J4TwistedTubs}
in order to handle stereo mini-jet cells of a cylindrical
drift chamber like JLC-CDC.
This new solid consists of three kinds of surfaces,
each of which is represented respectively by
{\tt J4HyperboloidalSurface},
{\tt J4FlatSurface}, and
{\tt J4TwistedSurface}.
These three surface classes correspond to
inner and outer hyperboloidal surfaces,
two end planes,
and two so-called twisted surfaces that make
slant and twisted $\phi$-boundaries, respectively.

There has been no Geant4 object for
the twisted surface.
In this paper, we have thus explained the algorithmic details
of our new surface class ({\tt J4TwistedSurface}) for it.
It should also be emphasized that
all of these three surface classes
are derived from a single base class called {\tt J4VSurface},
which greatly facilitated the distance and surface normal calculations.

We have implemented stereo cells with the new solid,
and tested them using geantinos and
Pythia events ($e^{+}e^{-}\rightarrow ZH$ at $\sqrt{s} = 350~{\rm GeV}$).
The stereo cells consumed 25\% more CPU time than
ordinary axial cells did.
We found this acceptable, 
considering the expected CPU time necessary for calorimeter simulation.

\section*{Acknowledgments}

The authors would like to thank A.~Miyamoto, T.~Aso, N.~Khalatyan, R.~Kuboshima,
and other members of the JLC-Software and JLC-CDC groups
for useful discussions and helps.
They are also grateful to Geant4 users group and developers.
In particular, a discussion with G. Cosmo was 
very useful for one of the authors.

\end{document}